# PreMevE Update: Forecasting Ultra-relativistic Electrons inside Earth's Outer Radiation Belt

Saurabh Sinha[*,1,2], Yue Chen[†,1], Youzuo Lin[1], and Rafael Pires de Lima[3]

[1]Los Alamos National Laboratory, Los Alamos, New Mexico, USA

[2]University of Oklahoma, Norman, OK, USA

[3]Geological Survey of Brazil, São Paulo, Brazil

[*]Correspondence to: saurabh.sinha@ou.edu

[†]Correspondence to: cheny@lanl.gov

**Key Points:**

- Machine-learning based PreMevE model is extended to predict ultra-relativistic electron flux distributions during MeV electron events

- This new PreMevE-2E model makes reliable 1- and 2- day ensemble forecasts of ≥2 MeV electrons inside Earth's outer radiation belt

- Nonlinear components play a major role in this new model at small L-shells (< ~4) in contrast to previous PreMevE for 1 MeV electrons

**Abstract.** Energetic electrons inside Earth's outer Van Allen belt pose a major radiation threat to space-borne electronics that often play vital roles in our modern society. Ultra-relativistic electrons with energies greater than or equal to two Megaelectron-volt (MeV) are of particular interest due to their high penetrating ability, and thus forecasting these ≥2 MeV electron levels has significant meaning to all space sectors. Here we update the latest development of the predictive model for MeV electrons inside the Earth's outer radiation belt. The new version, called PreMevE-2E, focuses on forecasting ultra-relativistic electron flux distributions across the outer radiation belt, with no need of in-situ measurements except for at the geosynchronous (GEO) orbit. Model inputs include precipitating electrons observed in low-Earth-orbits by NOAA satellites, upstream solar wind conditions (speeds and densities) from solar wind monitors, as well as ultra-relativistic electrons measured by one Los Alamos GEO satellite. We evaluated a total of 32 supervised machine learning models that fall into four different classes of linear and neural network architectures, and also successfully tested ensemble forecasting by using groups of top-performing models. All models are individually trained, validated, and tested by in-situ electron data from NASA's Van Allen Probes mission. It is shown that the final ensemble model generally outperforms individual models overs L-shells, and this PreMevE-2E model provides reliable and high-fidelity 25-hr (∼1-day) and 50-hr (∼2-day) forecasts with high mean performance efficiency values. Our results also suggest this new model is dominated by non-linear components at low L-shells (< ∼4) for ultra-relativistic electrons, which is different from the dominance of linear components at all L-shells for 1 MeV electrons as previously discovered.

# 1. Introduction

Since their discovery in 1958, energetic particles inside the Earth's Van Allen radiation belts have been one top concern for space operations, including the Apollo missions in early years of the Space Age. Our interests in these magnetically trapped electrons and protons have been repeatedly refreshed, and understanding of these belt particles has deepened continuously as observations accumulated over the past six decades. It is well recognized that these particles usually present a two-belt distribution—an inner belt in the region with equatorial distances (i.e., L-shells or simply L) within ~2-3 Earth radii and an outer belt with ~3 < L < 8 separated by the slot region in between. This general picture for the electron belts has been widely accepted by the aerospace industry, as specified by empirical models such as the AE8 (Vette, 1991). Starting from 1990, observations from the Combined Radiation Release and Effects Satellite—particularly the deep injection of Megaelectron-volt (MeV) electrons during the March 1991 event (Blake et al., 1992)—reignited the research interest in understanding the dynamics of outer belt electrons, whose intensities may vary up to several orders of magnitude during magnetic storms. Recently, the latest observations from Van Allen Probes (also called RBSP) again surprised the space community by showing the persistent absence of MeV electrons inside the inner belt (Fennell et al., 2015 and Claudepierre et al., 2015).

Indeed, for satellites operating in geosynchronous orbit (GEO), geosynchronous-transfer-orbit (GTO), medium- and high-earth-orbits (MEO and HEOs) with high apogees, energetic electrons inside the outer belt pose a major space radiation risk, not only in term of the ionizing dose, but also due to the deep dielectric charging and discharging phenomena (Reagan et al. 1983). When space systems are irradiated, some of the electrons are energetic enough to penetrate through satellite surfaces (e.g., the range of 2 MeV electrons inside

Aluminum is 178 mil), stop and bury themselves inside the dielectric materials of electronic parts on board. During major MeV electron events when electron intensities across the outer belt are greatly enhanced to sustaining high levels, these buried electrons accumulate faster than they dissipate, and thus build up high electric fields (called charging with the potential differences sometimes up to multiple kilovolts) in these regions, until eventually sudden intense breakdowns occur and the resulted discharging arcs may cause catastrophic failures to electronics and satellites.

Consequently, understanding and forecasting MeV electron events have been a central research topic for radiation belt studies. Recent mounting evidence, particularly from Van Allen Probes, has suggested that local wave-particle interactions play a critical role in energizing seed electrons to MeV and above in individual events, while the radial diffusion can also be important (e.g., Li and Hudson, 2019 and references therein). Based on this theoretic framework, a list of first-principle three-dimensional diffusive models have been developed and shown their successes as well as limits in describing MeV electron dynamics (see Chen et al., 2016 for a brief review), while a different approach has also been proposed and explored by using precipitating low-energy electrons observed in low-Earth-orbits (LEO) as a proxy for the wave-particle interactions. This new idea of predicting MeV electron events inside the outer belt was first presented by Chen et al. (2016), showing case a short interval for feasibility demonstration, and then Chen et al. (2019) successfully constructed the first PREdictive MEV Electron (PreMevE) model based on simple linear predictive filters. The follow-up study by Pires de Lima et al. (2020) (abbreviated to P2020 hereinafter) advanced the model to PreMevE 2.0, by fusing machine learning (ML) algorithms, which is able to reliably forecast 1 MeV electron distributions across the outer belt. This current study further expands forecasts to electrons with higher energies ≥ 2 MeV, the population with high beta ratio (velocity over light speed) values > ~0.98 and thus also called ultra-relativistic electrons in this work.

Another new component of this study is the utilizing of ensemble forecasting, a predictive skill that has been widely adopted in meteorology forecasts. Ensemble forecasting is a numerical method that uses multiple predictions from slightly different initial conditions, or different forecast models, to generate a broad sample of the possible future states of a dynamical system (Knipp, 2016). The instances of different conditions or different models are called "members." In the forecast cycle each member starts with a current state of the system based on a combination of observations and a background model, followed by a calculation of the system evolution over time. Outputs from the members are then combined and analyzed for trends and uncertainty ranges. Recently, ensemble methods have been applied to a list of space research, ranging from predicting new solar cycle and coronal mass ejections to magnetospheric reactions, to which a brief review was given by Knipp (2016) and references therein.

Purpose of this paper is to report how PreMevE has been upgraded to make predictions of ultra-relativistic electron flux distributions across the outer radiation belt. Still, with no requirement of in situ electron measurements except for at GEO, this unique model, named PreMevE-2E where E stands for both ensemble forecast and enhancement, has enhanced its capability to meet the predictive requirements for penetrating outer-belt electrons during the post-RBSP era. In the next section, data and parameters to be used for this study are briefly described, as well as the selected ML algorithms. Section 3 explains in details how the model is trained, validated and tested to forecast the distributions of >2 MeV electron integral fluxes, followed by the forecasts on 2 MeV electron differential fluxes in Section 4. This work is concluded by Section 5 with a summary of findings and possible future directions.

## 2. Data, Parameters, and Machine Learning Algorithms

Most data, model parameters and ML algorithms used in this work have been previously described in detail by Chen et al. (2019) and P2020, and here is a brief recap focusing on the differences. Ultra-relativistic electron flux distributions are from the in-situ observations made by Relativistic Electron-proton Telescope (REPT, Baker et al., 2012) experiment aboard RBSP-a spacecraft at $L \leq 6$, and by Energy Spectrometer for Particles (ESP, Meier et al., 1996) instrument carried by one Los Alamos National Laboratory (LANL) GEO satellite LANL-01A at $L = 6.6$. As presented in Figure 1A, integral fluxes of >2 MeV electrons are the target data set that is a function of L-shell over a 1289-day interval starting from 2013 February 20. These >2 MeV electron data are used for model training, validation and test, and are not needed as model input (except for at GEO) for making predictions.

Model input parameters include low-energy precipitating electrons measured by one NOAA Polar Operational Environmental Satellite (POES) NOAA-15 and upstream solar wind conditions over the same time interval. As shown in Figure 1, POES electron data used here are the same as in P2020, and thus the same nomenclature are adopted: E2, E3, and P6 refer to the count rates of >100, >300, and >1000 keV electrons from different POES channels, as shown in Panels B – D, respectively. Hereinafter, all electron intensities for the target, E2, E3, and P6 data are in logarithmic values unless being specified otherwise. Upstream solar wind conditions include the speeds that have been tested in P2020, and solar wind densities (SWD) as the new model input parameter. All electron intensities as well as solar wind parameters in Figure 1 are binned by 5 hr for the time, and electrons are also binned for L-shells with the size of 0.1. We standardized the solar wind speeds and densities by first subtracting their mean values and then dividing the results with the standard deviations. The same four supervised ML algorithms as tested in P2020 are used due to their previous success, including linear regression, feedforward neural networks (FNNs), long short-term memory (LSTM), and convolutional neural networks (CNNs). Briefly, linear regression models seek the optimized linear relationship between input parameters and targets. FNNs use layers of neurons to process inputs with linear transformations followed by nonlinear activation functions to optimize outputs. LSTM networks consist of connected memory cells that learn the sequential and temporal dynamics from the previous time steps to make predictions, and CNNs rely upon a convolution kernel to filter the data and explore the local patterns inside. All FNNs, LSTM and CNNs are trained with the objective to digest inputs and minimize a specified loss function. More details of these algorithms can be found in P2020 and references therein.

For model development, data in Figure 1 are split for training, validation, and test, with portions of 65% (~835 days), 14% (175), and 21% (267), respectively. Models are trained for each individual L-shell between 2.8 and 6 as well as at GEO (6.6) in the outer belt region, with the optimization goal of minimizing the root-mean-square error between the target values $y$ (electron fluxes in logarithm) and predicted values $f$. Different parameter combinations and temporal window sizes are tested for model inputs. We also compare the performance of models using different ML algorithms as in P2020. Model performance is gauged by Performance Efficiency (PE), which quantifies the accuracy of predictions by comparing to the variance of the target. Naming $y$ and $f$ both with size $M$, PE is defined as

$$PE = 1 - \frac{\sum_{j=1}^{M}(y_j - f_j)^2}{\sum_{j=1}^{M}(y_j - \bar{y})^2},$$ where $\bar{y}$ is the mean of $\mathbf{y}$. PE does not have a lower bound, and its perfect score is 1.0, meaning all predicted value perfectly match observed data, or that $\mathbf{f} = \mathbf{y}$.

**3. Forecasting >2 MeV Electron Flux Distributions**

Models in this section predict the integral fluxes of >2 MeV electrons. Key results are summarized in Tables 1 and 2, followed by detailed discussions. Table 1 lists PE values of all 32 models for 25 hr (or called 1-day) forecasts, and Table 2 are for 50 hr (2-day) forecasts. In each Table, there are eight input and window size combinations for each of the four ML algorithms, where the model names follow the convention of P2020. For example, FNN-64-32-elu are FNNs composed of two hidden layers—the first one has 64 neurons and the second has 32 neurons, and the neutrons use Exponential Linear Unit (ELU, Clevert et al., 2015) as the activation function; LSTM-128 models have one layer with 128 memory cells; and conv-64-32-relu are CNN models composed of two convolutional layers—the first one contains 64 kernels and the second contains 32 kernels, and the kernels use Rectified Linear Unit (ReLU, Hahnloser et al., 2000; Nair & Hinton, 2010) as an activation function. "Window size" refers to how many 5-hr time bins of input data are needed by the models. In the Input Parameters column, the dE2 refers to the normalized temporal derivatives of E2 fluxes, E246 indicates E2 fluxes at L = 4.6 are used as inputs for all L-shells, and SW and SWD refer to solar wind speeds and densities, respectively. The last row in each Table is for the ensemble model that will be discussed later in this section.

First, we examined the effects of model hyperparameters (i.e., input combinations and window sizes) as in Figure 2, which uses linear and LSTM models as the examples and plots the out-of-sample PE values (for validation and test data) as a function of L-shells. In Panel A, the general trend can be observed for linear models that PE increases with the increasing number of input parameters and window sizes. All curves have similar shapes with the highest PE at L~4.0 and decreasing in both directions, until PE values grow above 0.6 at GEO. The high PE values at GEO can be explained by the inclusion of >2 MeV electron fluxes in-situ measured by LANL-01A satellite, as previously seen in P2020. Here we highlight three examples: models 6 and 8 have different input parameters but the same window size, while models 7 and 8 have the same input parameters but different window sizes (see Table 1). It is seen that model 8 has the highest PE with SWD included in inputs. In Panel B, LSTM models have very different PE curves with large variations. Several LSTM models show a local minimum in PE with L-shell at ~4 and a plateau at L between 3.1-3.8. In addition, the inclusion of SWD to models 23 and 24 indeed decrease their PE at L ≤ 6 compared to those of model 22 (also see Table 1). PE values can drop below zero at small L < 3.0, particularly for the linear models, mainly due to the lack of training events on which we have more discussion later in this section. Therefore, hereinafter we confine our discussions on PE only for L ≥ 3.0.

To get an idea of model performance, we first inspected Table 1 for 1-day forecasts comparing models' mean PE values, which are averaged over all L-shells except for GEO for individual models. Based on the mean out-of-sample PE values for combined validation and test data, one can rank models' performance from high to low. For instance, in the linear category, model 8 is the top performer that has the highest mean PE of 0.523, followed by model 6 with a PE value of 0.509. For the top performer model 8, its out-of-sample PE at GEO is 0.629, also the highest in the category and thus in bold and underscored. Similarly, the top and second performers in other categories are picked out with their mean PE in bold font and underscored. In Table 1, mean PE values of the four top (second) performers are

0.523 (0.509) for linear, 0.553 (0.488) for FNN, 0.537 (0.521) for LSTM, and 0.479 (0.477) for CNN, respectively, while their PE values at GEO are 0.629 (0.625), 0.630 (0.603), 0.600 (0.581), and 0.598 (0.566) that are not necessarily the highest of each category. Note among the four top performers, only the linear model 8 has SWD in model inputs, while at GEO three out of the four models with the highest PE, i.e., models 8, 15 and 23, have SWD included.

Similarly, in Table 2 for 2-day forecasts, mean out-of-sample PE values for the four top (second) performers are 0.438 (0.431), 0.460 (0.416), 0.456 (0.451), and 0.423 (0.408), respectively, while their PE at GEO are 0.428 (0.431), 0.423 (0.419), 0.390 (0.384), and 0.402 (0.345) which are often not the highest in the category. For 2-day forecasts, SWD are not needed for the top four performers, while at GEO the only exception is the linear model 8. Therefore, unlike the important role of SW as demonstrated here and in P2020, SWD is not necessary for model input except for 1-day linear forecasts at GEO. Also, in both Tables 1 and 2, top FNN and LSTM models marginally outperform top linear models, suggesting the significance of non-linear component for >2 MeV electrons, in sharp contrast to P2020 models for 1 MeV electrons in which top linear ones always have the highest (or next to the highest) PE values. Additionally, PE values at GEO are ~0.1 higher than the mean PE at L≤6 for 1-day forecasts, while for 2-day forecasts PE values are slightly lower at GEO.

PE curves for the top two performers in each category for one- and two-day forecasts are further compared in Figure 3 as a function of L-shell. First, note in both panels there is no one individual model outperforms others over all L-shells. For example, linear models (i.e., the solid gray curves) have higher PE at L-shells above ~3.8, while the top FNN (red) and LSTM (brown) models perform better at small L-shells than the quickly degrading linear ones. Plus, the PE curves for the top linear model 6 of PreMevE 2.0 for 1 MeV electrons (see Tables 2 and 3 in P2020) are plotted in dashed gray for comparison. Taking the linear models for example, the linear ones of this new model in solid gray have higher PE at L-shells >4.5 for 1-day (>4.0 for 2-day) but lower PE at smaller L-shells.

An overview of 1-day forecasted flux distributions from the four top models are presented in Figure 4, showing similar dynamics compared to those observed in target data. Over the whole interval, most MeV electron events are captured well in terms of both intensities and L-shell ranges, e.g., the areas in red and yellow colors. Exceptions include the significant electron dropouts, e.g., the vertical blue strip on days ~ 1080 at L > 5, and the deep electron injections into small L-shells below 3.0. To highlight the differences between forecasts and target, error ratios for the four models are plotted in Figure 5, in which green color indicates perfect predictions while blue (red) means over-(under-) predictions. For example, in the validation and test periods, the lack of vertical red strips suggests the onsets of >2 MeV electron events are well predicted, while the vertical blue strips reflect the predicted high fluxes during dropouts, which is acceptable since this model aims to predict the enhancements of energetic electrons. Again, the reddish areas at small L-shells ~ 2.8 and 2.9 during the validation and test periods, particularly in Panel A, indicate models' under par performance in the area. This is due to the fact that at these low L-shells training data is dominated by background and the ML algorithms can learn only from the single major event starting on day ~758, while there are up to three events during the validation and test periods. To keep the comparison standardized, we chose not to modify the training set for L-shells 2.8 and 2.9, and hence predictions at these L-shells are not as good as others. For the same reason, we excluded these two L-shells and only counted 3.0 and higher to calculate the mean PE values over L-shells as in Tables 1 and 2.

Alternatively, models' performance on 1-day forecasts can also be examined from scatter plots of flux data points over the whole 1289-day interval, as shown in Figure 6 for the top four models. In each 2D histogram, the position of each pixel compares the predicted and target fluxes and the pixel color counts the occurrences over the interval. In each panel, the diagonal indicates a perfect match, and the dark gray (light gray) dashed lines on both sides mark error factor ratios of 3 (5) and 1/3 (1/5) between predicted and observed fluxes (original flux values not in logarithm). The majority of the points (in red) fall close to the diagonal and are well contained between the two factor-3 lines, particularly the points in the upper right quarter during MeV electron events. The two percentages in the lower right show how many data points fall with the two pairs of factor lines, and the red number in the second row is the correlation coefficient (CC) value. It is seen that all models have high CC values from 0.90 to 0.91, and 71 - 78% (87-90%) forecasts have error ratios within the factors of 3 (5). In addition to PE, all these numbers further quantify the good performance of the top four models.

The overview plot for 2-day forecasts is given by Figure 7 for the four top performers as identified in Table 2. Similar features can be seen here as in Figure 4, including the resemblance between forecasts and observations as well as the misses at low L-shells of 2.8 and 2.9. Recalling the L-shell dependent performance of models as shown in Figure 3, we decided to test ensemble forecasts for optimization, using a combination of linear and non-linear models. Indeed, as mentioned in Section 1, the ensemble forecast has been widely applied for weather forecasting (e.g., see the review by Cheung et al., 2001), and studies have shown that the ensemble mean can act as a non-linear filter with a skill statistically higher than any ensemble individual member (Toth and Kalnay, 1997).

We first tested 1-day forecasts using a small ensemble group. As shown in the last row of Table 1, ensemble members include linear model 8, FNN model 13, LSTM model 22 and CNN model 29, which are the top four performing models in each of the four categories. At each time step the ensemble prediction of electron fluxes at one L-shell is the median of all four member model outputs, and standard deviation of the outputs is the measure of uncertainty. One such example is shown in Figure 8, where at four individual L-shells the ensemble forecasts (in red) closely track the increments and decays of >2 MeV electron fluxes (black) observed during MeV electron events, and the gray strips in the background represent the uncertainties from this ensemble group. Similarly, another ensemble group was constructed for 2-day forecasts, with the same member models except for the linear one being replaced by model 8 (see the last row in Table 2), and the ensemble forecasts are also shown to closely trace observations at four individual L-shells as in Figure 9.

The plots in Figure 10 compare the observed and ensemble forecasted flux distributions over the whole interval. There are noticeable improvements, including the better predictions of low fluxes at L-shells ~3.5, e.g., the blue area centered on day 552 during the training in Panel B, and the deep injections to low L-shells during the validation and test periods when compared to the linear model in Figure 4B. In general, however, it is not easy to tell the difference just by eyeballing and comparing to distributions in Figures 4, 5 and 7.

Therefore, we again use the PE to quantify model performance, comparing the ensemble PE curves to those of group members as a function of L-shell in Figure 11. First, it is seen that in both panels, the ensemble PE curves (in red) almost always stay to the rightmost for all L-shells, including at GEO, when compared to PE curves from four member models. The outperformance of ensemble models is consistent with previous results from other fields, and justifies the usage of ensemble forecasting in this new model. Second, when compared to the PE curves in dashed gray from the linear model of PreMevE 2.0, our ensemble forecasts either have at least comparable performance in Panel A for 1-day or have even better performance in Panel B for 2-day, in particular at medium or high L-shells. Looking back to Tables 1 and 2, the ensemble models have a mean PE value of 0.612 for 1-day and 0.521 for 2-day at L ≤ 6, and 0.677 and 0.572 at GEO, and all these values are significantly higher than those from individual top performer models. Therefore, our test demonstrates the advantage of ensemble forecasting, and thus this new model is named PreMevE-2E for its adoption of ensemble forecasting to enhance prediction capability. To put this model's performance into context, the operational Relativistic Electron Forecast Model (REFM) at NOAA has PE values of 0.72 and 0.49 at GEO for 1- and 2-day predictions for daily averaged fluence of >2 MeV electrons, and our model has PE values of ~0.68 and ~0.57 at GEO for 1- and 2-day forecasts of >2 MeV electron fluxes with 5 hr time resolution and additionally has similar predictive performance in the heart region of the outer belt (3 ≤ L ≤ 6).

Note here we only tested with a small ensemble group, and obviously there are other possible options. For example, the ensemble group may include more members, and not necessarily same number of models from all categories. Besides, particularly at GEO the ensemble group can be different, for instance by selecting the models with the highest out-of-sample PE values at GEO. All these require further extensive tests, which we leave to future study.

## 4. Forecasting 2 MeV Electron Flux Distributions

This section explains how PreMevE-2E predicts differential flux distributions of 2 MeV electrons. The methodology is identical to those used for >2 MeV electrons as described in Section 3, and here we summarize the results. First, the effects of model hyperparameters are tested and the mean PE values for individual models are presented in Tables 3 and 4 for 1- and 2-day forecasts, respectively. In Table 3, mean PE values of the four top (second) performers are 0.600 (0.590) for linear, 0.549 (0.548) for FNN, 0.549 (0.533) for LSTM, and 0.525 (0.518) for CNN, respectively, while their PE values at GEO are 0.566 (0.568), 0.461 (0.535), 0.509 (0.539), and 0.459 (0.437) that are lower than the highest for each category. In Table 4, mean PE values of the four top (second) performers are 0.512 (0.506) for linear, 0.474 (0.461) for FNN, 0.438 (0.435) for LSTM, and 0.439 (0.425) for CNN, respectively, while their PE values at GEO are 0.234 (0.244), 0.186 (0.105), 0.138 (0.106), and 0.102 (0.125) that are often not even close to the highest for each category. It is interesting to notice that the top (and second) linear models have higher mean PE than all the remaining top performers for both 1- and 2-day forecasts. Based on the rank of mean PE values, in the last row of both Tables, ensemble forecasting models are constructed including the top performers from each of the four categories.

The overview plots in Figure 12 compare the observed and ensemble forecasted flux distributions over the whole interval. The similarity between 1-day ensemble forecasts (Panel B) and target distributions (Panel A) is impressive, but the vertical red strips in the error ratio distribution (Panel C) at L > ~4 also suggest the forecasts often miss the very beginning of the onsets of MeV electron events. Similar features are seen in Panels D and E for 2-day ensemble forecasts.

In Figure 13, model performance is quantified by comparing the ensemble PE curves to those of group members as a function of L-shell. First, as seen in Figure 11, in both panels the ensemble PE curves (in red) almost always stay to the rightmost for all Lshells, except for at GEO for 2-day, when compared to the PE curves from four member models. Therefore, the outperformance of ensemble models is confirmed for 2 MeV electrons. Second, when compared to the PE curves in dashed gray from the linear model of PreMevE 2.0, our ensemble forecasts have comparable (Panel B) or even better (A) performance in average but not at GEO. From the last rows of Tables 3 and 4, the ensemble models have a mean PE value of 0.624 for 1-day and 0.521 for 2-day at $L \leq 6$, and 0.564 and 0.186 at GEO, and all these mean PE values are higher than those from individual top performer models but not at GEO. To show more details, for 1-day forecasts, Figure 14 shows at four individual L-shells the ensemble forecasts (in red) closely track the ups and downs of 2 MeV electron fluxes (black), and similarly Figure 15 shows 2-day forecast results. It is noticeable that in Panel D the 2-day forecasts at GEO often have values much lower than those observed peak flux values, in particular during the several major events, which may explain the low PE value of 0.186 at GEO.

There are two motivations for us to test predicting differential flux distributions of 2 MeV electrons. One is to have a counterpart so as to compare with PreMevE 2.0 model for 1 MeV electrons, and it turns out the two models have very similar performance in term of PE (except for 2-day at GEO). The other motivation is that, with both integral and differential fluxes available, one can further determine a single-parameter energy spectrum shape (e.g., in an exponential form) that can be helpful to quantify radiation effects (e.g., ionizing doses) with a given satellite geometry and shielding design.

## 5. Summary and Conclusions

Using electron data from NASA's RBSP mission, we have trained, evaluated and tested a set of supervised machine learning models to forecast ≥2 MeV electron fluxes. After evaluating the performance of these models, ensemble forecasting has proven overall to perform better than any individual model in different categories. After the completeness of training, our model has demonstrated to make reliable forecasts with no more need of in-situ electron measurements from RBSP.

This new PreMevE-2E model can well predict dynamic distributions of ultra-relativistic electrons, with measurement inputs that are made available from satellites operating in LEO, GEO, and at the Lagrangian 1 point of the Sun-Earth system. In this work we have evaluated and tested that: 1) the effects of different parameter combinations, including solar wind densities, as well as the window sizes for model performance; 2) four categories of linear and neural network models; and 3) the adoption of ensemble forecasting. PreMevE-2E has enhanced its forecasting capability by extending to the ultra-relativistic electron energy range. Model predictions over a 14 months out-of-sample period demonstrate that this model provides high-fidelity 1-day (2-day) forecasts of ≥2 MeV electron flux distributions: the mean PE values are above 0.61 (0.52) for both integral and differential fluxes across L-shells from 3 to 6; at GEO, model PE values are ~0.68 and ~0.57 for 1- and 2-day forecasts of >2 MeV integral fluxes, and ~0.56 and ~0.19 for the differential fluxes of 2 MeV electrons. Therefore, we believe this newly updated PreMevE-2E model lays down another step stone towards the full preparedness for severe MeV electron events in the future.

**Acknowledgements**

The authors declare no conflicts of interest. We gratefully acknowledge the support of NASA Heliophysics Space Weather Operations to Research Program (18-HSWO2R18-0006), the NASA Heliophysics Guest Investigators program (14-GIVABR14_2-0028), and LANL internal funding. We want to acknowledge the PIs and instrument teams of NOAA POES SEM2 and RBSP REPT for providing measurements and allowing us to use their data. Thanks to CDAWeb for providing OMNI data. RBSP and POES data used in this work were downloadable from the missions' public data websites (https://www.rbsp-ect.lanl.gov and http://www.ngdc.noaa.gov), while LANL-01A electron data can be found in ASCII format at https://osf.io/6xzgw/.
18

**Tables**

**Table 1: Performance of models in four categories (>2 MeV) for 1-day (25 hr) forecasts.** Among the eight models in each category, the top performer—ranked by the out-of-sample performance efficiency (PE) values in the (PE val + test) column—has its model number in bold font and underscored, and the second performer has its number in bold. In the last column for PE at GEO for validation and test data sets, the highest PE value for each category is also in bold and underscored, and it may not be the same as the one from the top performer (which always has its GEO PE value underscored). The last row (model 33) is for the ensemble model PE values.

| Index | Models | Window size | Input Parameters | PE train | PE validation | PE test | PE val + test | PE all | PE GEO val+test |
|---|---|---|---|---|---|---|---|---|---|
| 1 | LinearReg | 4 | E2+E3+P6+SW | 0.712 | 0.108 | 0.454 | 0.414 | 0.707 | 0.621 |
| 2 | LinearReg | 16 | E2+E3+P6+SW | 0.742 | 0.194 | 0.509 | 0.470 | 0.736 | 0.623 |
| 3 | LinearReg | 4 | E2+E3+P6+SW+dE2 | 0.714 | 0.112 | 0.461 | 0.420 | 0.709 | 0.622 |
| 4 | LinearReg | 16 | E2+E3+P6+SW+dE2 | 0.747 | 0.197 | 0.523 | 0.479 | 0.741 | 0.625 |
| 5 | LinearReg | 4 | E2+E3+P6+SW+dE2+E246 | 0.736 | 0.188 | 0.486 | 0.456 | 0.731 | 0.622 |
| 6 | LinearReg | 16 | E2+E3+P6+SW+dE2+E246 | 0.763 | 0.255 | 0.548 | 0.509 | 0.757 | 0.625 |
| 7 | LinearReg | 4 | E2+E3+P6+SW+SWD+dE2+E246 | 0.741 | 0.193 | 0.502 | 0.466 | 0.736 | 0.627 |
| **8** | LinearReg | 16 | E2+E3+P6+SW+SWD+dE2+E246 | 0.770 | 0.266 | 0.568 | <u>0.523</u> | 0.764 | <u>0.629</u> |
| 9 | FNN-64-32-elu | 4 | E2+E3+P6+SW | 0.686 | 0.202 | 0.463 | 0.426 | 0.690 | 0.631 |
| 10 | FNN-64-32-elu | 16 | E2+E3+P6+SW | 0.699 | 0.331 | 0.459 | 0.460 | 0.704 | 0.620 |
| 11 | FNN-64-32-elu | 4 | E2+E3+P6+SW+dE2 | 0.644 | 0.216 | 0.403 | 0.395 | 0.658 | 0.630 |
| 12 | FNN-64-32-elu | 16 | E2+E3+P6+SW+dE2 | 0.716 | 0.319 | 0.511 | 0.488 | 0.720 | 0.603 |
| **13** | FNN-64-32-elu | 4 | E2+E3+P6+SW+dE2+E246 | 0.766 | 0.404 | 0.566 | <u>0.553</u> | 0.765 | <u>0.630</u> |
| 14 | FNN-64-32-elu | 16 | E2+E3+P6+SW+dE2+E246 | 0.704 | 0.200 | 0.441 | 0.408 | 0.699 | 0.624 |
| 15 | FNN-64-32-elu | 4 | E2+E3+P6+SW+SWD+dE2+E246 | 0.713 | 0.266 | 0.456 | 0.446 | 0.713 | <u>0.646</u> |
| 16 | FNN-64-32-elu | 16 | E2+E3+P6+SW+SWD+dE2+E246 | 0.715 | 0.195 | 0.392 | 0.384 | 0.703 | 0.621 |
| 17 | LSTM-128 | 4 | E2+E3+P6+SW | 0.662 | 0.208 | 0.445 | 0.414 | 0.673 | 0.527 |
| 18 | LSTM-128 | 16 | E2+E3+P6+SW | 0.750 | 0.366 | 0.537 | 0.521 | 0.747 | 0.581 |
| 19 | LSTM-128 | 4 | E2+E3+P6+SW+dE2 | 0.665 | 0.198 | 0.440 | 0.410 | 0.675 | 0.538 |
| 20 | LSTM-128 | 16 | E2+E3+P6+SW+dE2 | 0.740 | 0.287 | 0.526 | 0.489 | 0.737 | 0.588 |
| 21 | LSTM-128 | 4 | E2+E3+P6+SW+dE2+E246 | 0.700 | 0.282 | 0.472 | 0.459 | 0.706 | 0.535 |
| **22** | LSTM-128 | 16 | E2+E3+P6+SW+dE2+E246 | 0.781 | 0.401 | 0.545 | <u>0.537</u> | 0.771 | <u>0.600</u> |
| 23 | LSTM-128 | 4 | E2+E3+P6+SW+SWD+dE2+E246 | 0.671 | 0.140 | 0.387 | 0.365 | 0.674 | <u>0.648</u> |
| 24 | LSTM-128 | 16 | E2+E3+P6+SW+SWD+dE2+E246 | 0.799 | 0.348 | 0.507 | 0.499 | 0.777 | 0.571 |
| 25 | Conv-64-32-relu | 4 | E2+E3+P6+SW | 0.702 | 0.289 | 0.462 | 0.453 | 0.705 | 0.593 |
| 26 | Conv-64-32-relu | 16 | E2+E3+P6+SW | -0.178 | -3.765 | -2.170 | -2.333 | -0.341 | -0.002 |
| 27 | Conv-64-32-relu | 4 | E2+E3+P6+SW+dE2 | 0.710 | 0.292 | 0.477 | 0.462 | 0.711 | 0.596 |
| 28 | Conv-64-32-relu | 16 | E2+E3+P6+SW+dE2 | 0.186 | -2.251 | -1.138 | -1.268 | 0.078 | -0.081 |
| **29** | Conv-64-32-relu | 4 | E2+E3+P6+SW+dE2+E246 | 0.719 | 0.324 | 0.480 | <u>0.479</u> | 0.722 | <u>0.598</u> |
| 30 | Conv-64-32-relu | 16 | E2+E3+P6+SW+dE2+E246 | 0.110 | -2.382 | -1.334 | -1.398 | 0.006 | -0.168 |
| 31 | Conv-64-32-relu | 4 | E2+E3+P6+SW+SWD+dE2+E246 | 0.749 | 0.285 | 0.497 | 0.477 | 0.742 | 0.566 |
| 32 | Conv-64-32-relu | 16 | E2+E3+P6+SW+SWD+dE2+E246 | 0.065 | -2.861 | -1.636 | -1.733 | -0.080 | 0.074 |
| 33 | Ensemble: models 8 + 13 + 22 + 29 | | | 0.782 | 0.393 | 0.625 | 0.612 | 0.783 | 0.677 |

**Table 2: Performance of models in four categories (>2 MeV) for 2-day (50 hr) forecasts.** In the same format as Table 1.

| Index | Models | Window size | Input Parameters | PE train | PE validation | PE test | PE val + test | PE all | PE GEO val+test |
|---|---|---|---|---|---|---|---|---|---|
| 1 | LinearReg | 4 | E2+E3+P6+SW | 0.675 | 0.049 | 0.381 | 0.352 | 0.671 | 0.417 |
| 2 | LinearReg | 16 | E2+E3+P6+SW | 0.702 | 0.120 | 0.433 | 0.400 | 0.696 | 0.421 |
| 3 | LinearReg | 4 | E2+E3+P6+SW+dE2 | 0.678 | 0.055 | 0.390 | 0.358 | 0.674 | 0.422 |
| 4 | LinearReg | 16 | E2+E3+P6+SW+dE2 | 0.707 | 0.127 | 0.444 | 0.409 | 0.701 | 0.427 |
| 5 | LinearReg | 4 | E2+E3+P6+SW+dE2+E246 | 0.701 | 0.128 | 0.411 | 0.391 | 0.695 | 0.422 |
| **6** | LinearReg | 16 | E2+E3+P6+SW+dE2+E246 | 0.727 | 0.189 | 0.468 | **0.438** | 0.720 | **0.428** |
| 7 | LinearReg | 4 | E2+E3+P6+SW+SWD+dE2+E246 | 0.703 | 0.141 | 0.420 | 0.399 | 0.697 | 0.429 |
| 8 | LinearReg | 16 | E2+E3+P6+SW+SWD+dE2+E246 | 0.606 | 0.450 | 0.414 | 0.431 | 0.577 | **0.431** |
| 9 | FNN-64-32-elu | 4 | E2+E3+P6+SW | 0.658 | 0.140 | 0.407 | 0.372 | 0.661 | 0.428 |
| 10 | FNN-64-32-elu | 16 | E2+E3+P6+SW | 0.669 | 0.251 | 0.392 | 0.393 | 0.671 | 0.403 |
| 11 | FNN-64-32-elu | 4 | E2+E3+P6+SW+dE2 | 0.624 | 0.174 | 0.367 | 0.360 | 0.638 | **0.438** |
| 12 | FNN-64-32-elu | 16 | E2+E3+P6+SW+dE2 | 0.686 | 0.236 | 0.436 | 0.416 | 0.687 | 0.419 |
| **13** | FNN-64-32-elu | 4 | E2+E3+P6+SW+dE2+E246 | 0.715 | 0.315 | 0.460 | **0.460** | 0.715 | **0.423** |
| 14 | FNN-64-32-elu | 16 | E2+E3+P6+SW+dE2+E246 | 0.670 | 0.109 | 0.373 | 0.340 | 0.664 | 0.433 |
| 15 | FNN-64-32-elu | 4 | E2+E3+P6+SW+SWD+dE2+E246 | 0.679 | 0.208 | 0.388 | 0.387 | 0.679 | 0.425 |
| 16 | FNN-64-32-elu | 16 | E2+E3+P6+SW+SWD+dE2+E246 | 0.681 | 0.133 | 0.322 | 0.322 | 0.668 | 0.385 |
| 17 | LSTM-128 | 4 | E2+E3+P6+SW | 0.634 | 0.141 | 0.386 | 0.359 | 0.644 | **0.425** |
| 18 | LSTM-128 | 16 | E2+E3+P6+SW | 0.711 | 0.290 | 0.461 | 0.451 | 0.708 | 0.384 |
| 19 | LSTM-128 | 4 | E2+E3+P6+SW+dE2 | 0.640 | 0.139 | 0.387 | 0.360 | 0.649 | 0.420 |
| 20 | LSTM-128 | 16 | E2+E3+P6+SW+dE2 | 0.706 | 0.238 | 0.453 | 0.428 | 0.701 | 0.360 |
| 21 | LSTM-128 | 4 | E2+E3+P6+SW+dE2+E246 | 0.668 | 0.193 | 0.394 | 0.385 | 0.672 | 0.366 |
| **22** | LSTM-128 | 16 | E2+E3+P6+SW+dE2+E246 | 0.739 | 0.307 | 0.457 | **0.456** | 0.729 | **0.390** |
| 23 | LSTM-128 | 4 | E2+E3+P6+SW+SWD+dE2+E246 | 0.644 | 0.110 | 0.342 | 0.328 | 0.647 | 0.418 |
| 24 | LSTM-128 | 16 | E2+E3+P6+SW+SWD+dE2+E246 | 0.743 | 0.252 | 0.405 | 0.407 | 0.723 | 0.335 |
| 25 | Conv-64-32-relu | 4 | E2+E3+P6+SW | 0.676 | 0.227 | 0.396 | 0.394 | 0.676 | 0.403 |
| 26 | Conv-64-32-relu | 16 | E2+E3+P6+SW | -0.115 | -3.656 | -2.088 | -2.227 | -0.283 | 0.048 |
| 27 | Conv-64-32-relu | 4 | E2+E3+P6+SW+dE2 | 0.684 | 0.237 | 0.407 | 0.404 | 0.683 | **0.403** |
| 28 | Conv-64-32-relu | 16 | E2+E3+P6+SW+dE2 | 0.209 | -2.242 | -1.139 | -1.254 | 0.094 | -1.397 |
| **29** | Conv-64-32-relu | 4 | E2+E3+P6+SW+dE2+E246 | 0.699 | 0.269 | 0.415 | **0.423** | 0.698 | **0.402** |
| 30 | Conv-64-32-relu | 16 | E2+E3+P6+SW+dE2+E246 | 0.181 | -2.220 | -1.240 | -1.287 | 0.070 | -0.105 |
| 31 | Conv-64-32-relu | 4 | E2+E3+P6+SW+SWD+dE2+E246 | 0.711 | 0.236 | 0.411 | 0.408 | 0.703 | 0.345 |
| 32 | Conv-64-32-relu | 16 | E2+E3+P6+SW+SWD+dE2+E246 | 0.125 | -2.594 | -1.530 | -1.588 | -0.020 | -0.328 |
| 33 | Ensemble: models 6 + 13 + 22 + 29 | | | 0.738 | 0.299 | 0.532 | 0.521 | 0.738 | 0.572 |

**Table 3: Performance of models in four categories (2 MeV) for 1-day (25 hr) forecasts.** In the same format as Table 1.

| Index | Models | Window size | Input Parameters | PE train | PE validation | PE test | PE val + test | PE all | PE GEO val+test |
|---|---|---|---|---|---|---|---|---|---|
| 1 | LinearReg | 4 | E2+E3+P6+SW | 0.746 | 0.358 | 0.591 | 0.538 | 0.744 | 0.549 |
| 2 | LinearReg | 16 | E2+E3+P6+SW | 0.769 | 0.427 | 0.628 | 0.583 | 0.768 | 0.561 |
| 3 | LinearReg | 4 | E2+E3+P6+SW+dE2 | 0.748 | 0.363 | 0.596 | 0.543 | 0.747 | 0.554 |
| 4 | LinearReg | 16 | E2+E3+P6+SW+dE2 | 0.773 | 0.432 | 0.637 | 0.590 | 0.772 | 0.568 |
| 5 | LinearReg | 4 | E2+E3+P6+SW+dE2+E246 | 0.761 | 0.383 | 0.594 | 0.550 | 0.755 | 0.553 |
| 6 | LinearReg | 16 | E2+E3+P6+SW+dE2+E246 | 0.781 | 0.440 | 0.632 | 0.590 | 0.776 | <u>0.568</u> |
| 7 | LinearReg | 4 | E2+E3+P6+SW+SWD+dE2+E246 | 0.768 | 0.385 | 0.595 | 0.551 | 0.759 | 0.555 |
| <u>8</u> | LinearReg | 16 | E2+E3+P6+SW+SWD+dE2+E246 | 0.792 | 0.450 | 0.645 | <u>0.600</u> | 0.784 | <u>0.566</u> |
| 9 | FNN-64-32-elu | 4 | E2+E3+P6+SW | 0.763 | 0.396 | 0.582 | 0.548 | 0.756 | 0.535 |
| 10 | FNN-64-32-elu | 16 | E2+E3+P6+SW | 0.721 | 0.298 | 0.481 | 0.451 | 0.710 | 0.471 |
| 11 | FNN-64-32-elu | 4 | E2+E3+P6+SW+dE2 | 0.679 | 0.105 | 0.411 | 0.344 | 0.663 | <u>0.582</u> |
| 12 | FNN-64-32-elu | 16 | E2+E3+P6+SW+dE2 | 0.700 | 0.238 | 0.476 | 0.426 | 0.693 | 0.505 |
| <u>13</u> | FNN-64-32-elu | 4 | E2+E3+P6+SW+dE2+E246 | 0.769 | 0.417 | 0.571 | <u>0.549</u> | 0.760 | <u>0.461</u> |
| 14 | FNN-64-32-elu | 16 | E2+E3+P6+SW+dE2+E246 | 0.727 | 0.249 | 0.455 | 0.422 | 0.708 | -0.735 |
| 15 | FNN-64-32-elu | 4 | E2+E3+P6+SW+SWD+dE2+E246 | 0.782 | 0.389 | 0.554 | 0.529 | 0.762 | 0.572 |
| 16 | FNN-64-32-elu | 16 | E2+E3+P6+SW+SWD+dE2+E246 | 0.730 | 0.168 | 0.418 | 0.369 | 0.695 | 0.487 |
| 17 | LSTM-128 | 4 | E2+E3+P6+SW | 0.705 | 0.295 | 0.493 | 0.456 | 0.702 | <u>0.578</u> |
| 18 | LSTM-128 | 16 | E2+E3+P6+SW | 0.751 | 0.387 | 0.539 | 0.518 | 0.742 | 0.539 |
| 19 | LSTM-128 | 4 | E2+E3+P6+SW+dE2 | 0.713 | 0.295 | 0.503 | 0.463 | 0.708 | 0.551 |
| 20 | LSTM-128 | 16 | E2+E3+P6+SW+dE2 | 0.744 | 0.360 | 0.519 | 0.496 | 0.731 | 0.509 |
| 21 | LSTM-128 | 4 | E2+E3+P6+SW+dE2+E246 | 0.757 | 0.362 | 0.539 | 0.511 | 0.746 | 0.577 |
| 22 | LSTM-128 | 16 | E2+E3+P6+SW+dE2+E246 | 0.791 | 0.423 | 0.525 | 0.525 | 0.764 | 0.516 |
| 23 | LSTM-128 | 4 | E2+E3+P6+SW+SWD+dE2+E246 | 0.782 | 0.395 | 0.556 | 0.533 | 0.764 | 0.539 |
| <u>24</u> | LSTM-128 | 16 | E2+E3+P6+SW+SWD+dE2+E246 | 0.836 | 0.453 | 0.551 | <u>0.549</u> | 0.795 | <u>0.509</u> |
| 25 | Conv-64-32-relu | 4 | E2+E3+P6+SW | 0.762 | 0.309 | 0.548 | 0.496 | 0.744 | 0.437 |
| 26 | Conv-64-32-relu | 16 | E2+E3+P6+SW | 0.602 | -0.596 | -0.096 | -0.186 | 0.494 | -0.549 |
| 27 | Conv-64-32-relu | 4 | E2+E3+P6+SW+dE2 | 0.770 | 0.340 | 0.566 | 0.518 | 0.754 | <u>0.437</u> |
| 28 | Conv-64-32-relu | 16 | E2+E3+P6+SW+dE2 | 0.637 | -0.467 | 0.031 | -0.057 | 0.545 | -0.490 |
| <u>29</u> | Conv-64-32-relu | 4 | E2+E3+P6+SW+dE2+E246 | 0.782 | 0.373 | 0.556 | <u>0.525</u> | 0.762 | <u>0.459</u> |
| 30 | Conv-64-32-relu | 16 | E2+E3+P6+SW+dE2+E246 | 0.638 | -0.422 | -0.050 | -0.090 | 0.533 | -0.898 |
| 31 | Conv-64-32-relu | 4 | E2+E3+P6+SW+SWD+dE2+E246 | 0.801 | 0.329 | 0.540 | 0.500 | 0.766 | 0.430 |
| 32 | Conv-64-32-relu | 16 | E2+E3+P6+SW+SWD+dE2+E246 | 0.670 | -0.537 | -0.086 | -0.170 | 0.529 | -0.631 |
| 33 | Ensemble: models 8 + 13 + 24 + 29 | | | 0.810 | 0.476 | 0.640 | 0.624 | 0.796 | 0.564 |

**Table 4: Performance of models in four categories (2 MeV) for 2-day (50 hr) forecasts.** In the same format as Table 1.

| Index | Models | Window size | Input Parameters | PE train | PE validation | PE test | PE val + test | PE all | PE GEO val+test |
|---|---|---|---|---|---|---|---|---|---|
| 1 | LinearReg | 4 | E2+E3+P6+SW | 0.701 | 0.288 | 0.500 | 0.461 | 0.700 | 0.186 |
| 2 | LinearReg | 16 | E2+E3+P6+SW | 0.721 | 0.339 | 0.535 | 0.497 | 0.720 | 0.227 |
| 3 | LinearReg | 4 | E2+E3+P6+SW+dE2 | 0.703 | 0.294 | 0.506 | 0.466 | 0.702 | 0.200 |
| 4 | LinearReg | 16 | E2+E3+P6+SW+dE2 | 0.725 | 0.349 | 0.544 | 0.506 | 0.724 | 0.244 |
| 5 | LinearReg | 4 | E2+E3+P6+SW+dE2+E246 | 0.717 | 0.309 | 0.500 | 0.469 | 0.710 | 0.198 |
| 6 | LinearReg | 16 | E2+E3+P6+SW+dE2+E246 | 0.735 | 0.354 | 0.534 | 0.502 | 0.729 | <u>0.244</u> |
| 7 | LinearReg | 4 | E2+E3+P6+SW+SWD+dE2+E246 | 0.720 | 0.323 | 0.504 | 0.475 | 0.714 | 0.212 |
| <u>8</u> | LinearReg | 16 | E2+E3+P6+SW+SWD+dE2+E246 | 0.743 | 0.364 | 0.546 | <u>0.512</u> | 0.735 | <u>0.234</u> |
| <u>9</u> | FNN-64-32-elu | 4 | E2+E3+P6+SW | 0.718 | 0.333 | 0.494 | <u>0.474</u> | 0.713 | <u>0.186</u> |
| 10 | FNN-64-32-elu | 16 | E2+E3+P6+SW | 0.669 | 0.212 | 0.395 | 0.370 | 0.661 | 0.222 |
| 11 | FNN-64-32-elu | 4 | E2+E3+P6+SW+dE2 | 0.642 | 0.041 | 0.348 | 0.286 | 0.627 | <u>0.258</u> |
| 12 | FNN-64-32-elu | 16 | E2+E3+P6+SW+dE2 | 0.661 | 0.143 | 0.394 | 0.346 | 0.650 | -0.044 |
| 13 | FNN-64-32-elu | 4 | E2+E3+P6+SW+dE2+E246 | 0.718 | 0.332 | 0.472 | 0.461 | 0.710 | 0.105 |
| 14 | FNN-64-32-elu | 16 | E2+E3+P6+SW+dE2+E246 | 0.685 | 0.153 | 0.373 | 0.341 | 0.665 | -0.891 |
| 15 | FNN-64-32-elu | 4 | E2+E3+P6+SW+SWD+dE2+E246 | 0.725 | 0.306 | 0.446 | 0.436 | 0.706 | 0.081 |
| 16 | FNN-64-32-elu | 16 | E2+E3+P6+SW+SWD+dE2+E246 | 0.681 | 0.087 | 0.327 | 0.289 | 0.647 | 0.146 |
| 17 | LSTM-128 | 4 | E2+E3+P6+SW | 0.667 | 0.197 | 0.407 | 0.373 | 0.660 | 0.191 |
| 18 | LSTM-128 | 16 | E2+E3+P6+SW | 0.699 | 0.270 | 0.435 | 0.416 | 0.687 | 0.204 |
| 19 | LSTM-128 | 4 | E2+E3+P6+SW+dE2 | 0.673 | 0.191 | 0.409 | 0.373 | 0.663 | <u>0.265</u> |
| 20 | LSTM-128 | 16 | E2+E3+P6+SW+dE2 | 0.689 | 0.237 | 0.435 | 0.404 | 0.678 | 0.224 |
| 21 | LSTM-128 | 4 | E2+E3+P6+SW+dE2+E246 | 0.702 | 0.261 | 0.439 | 0.418 | 0.691 | 0.201 |
| <u>22</u> | LSTM-128 | 16 | E2+E3+P6+SW+dE2+E246 | 0.735 | 0.322 | 0.440 | <u>0.438</u> | 0.712 | <u>0.138</u> |
| 23 | LSTM-128 | 4 | E2+E3+P6+SW+SWD+dE2+E246 | 0.715 | 0.271 | 0.426 | 0.413 | 0.696 | 0.247 |
| 24 | LSTM-128 | 16 | E2+E3+P6+SW+SWD+dE2+E246 | 0.772 | 0.327 | 0.436 | 0.435 | 0.730 | 0.106 |
| 25 | Conv-64-32-relu | 4 | E2+E3+P6+SW | 0.716 | 0.248 | 0.463 | 0.425 | 0.700 | <u>0.125</u> |
| 26 | Conv-64-32-relu | 16 | E2+E3+P6+SW | 0.562 | -0.694 | -0.195 | -0.279 | 0.448 | -0.332 |
| 27 | Conv-64-32-relu | 4 | E2+E3+P6+SW+dE2 | 0.720 | 0.236 | 0.451 | 0.413 | 0.699 | 0.105 |
| 28 | Conv-64-32-relu | 16 | E2+E3+P6+SW+dE2 | 0.670 | -0.359 | 0.101 | 0.031 | 0.582 | -1.124 |
| <u>29</u> | Conv-64-32-relu | 4 | E2+E3+P6+SW+dE2+E246 | 0.740 | 0.294 | 0.456 | <u>0.439</u> | 0.717 | <u>0.102</u> |
| 30 | Conv-64-32-relu | 16 | E2+E3+P6+SW+dE2+E246 | 0.584 | -0.605 | -0.203 | -0.245 | 0.464 | -0.980 |
| 31 | Conv-64-32-relu | 4 | E2+E3+P6+SW+SWD+dE2+E246 | 0.745 | 0.219 | 0.425 | 0.393 | 0.708 | 0.076 |
| 32 | Conv-64-32-relu | 16 | E2+E3+P6+SW+SWD+dE2+E246 | 0.680 | -0.530 | -0.117 | -0.173 | 0.532 | -1.216 |
| 33 | Ensemble: models 8 + 9 + 22 + 29 | | | 0.743 | 0.361 | 0.543 | 0.521 | 0.736 | 0.186 |

**Figures**

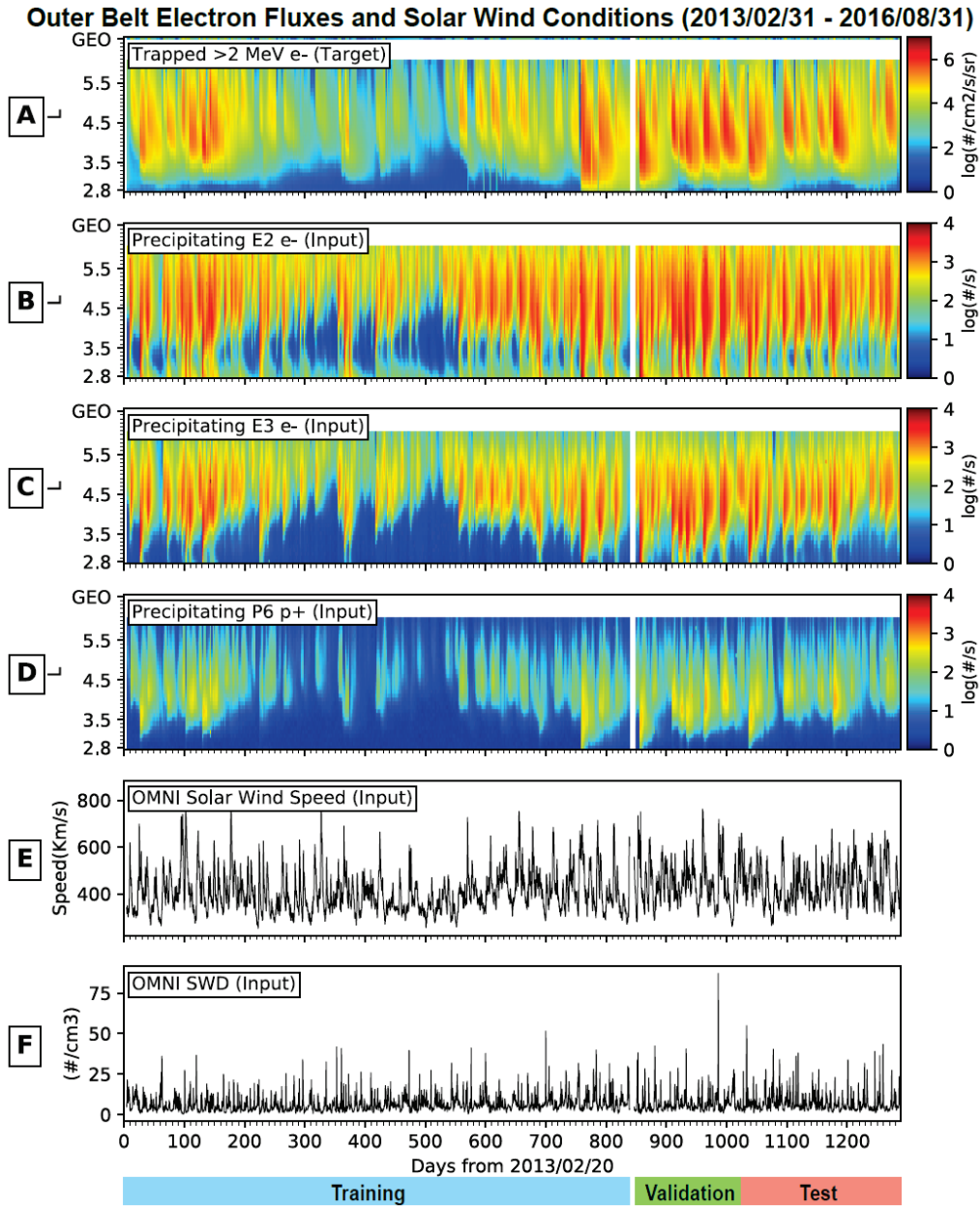

**Figure 1 Overview of electron observations and solar wind conditions.** All panels present for the same 1289-day interval starting from 2013/02/20. **A)** Flux distributions of >2 MeV electrons, the variable to be forecasted (i.e., targets). **B to D)** Count rates of precipitating electrons measured by NOAA-15 in LEO, for E2, E3, and P6 channels, respectively. **E)** Solar wind speeds measured upstream of the magnetosphere from the OMNI data set. **F)** Solar wind densities. Data in Panels **B** to **F** serve as model inputs, i.e., predictors. The bottom color bars indicate the portions of data used for training, validation, and test.

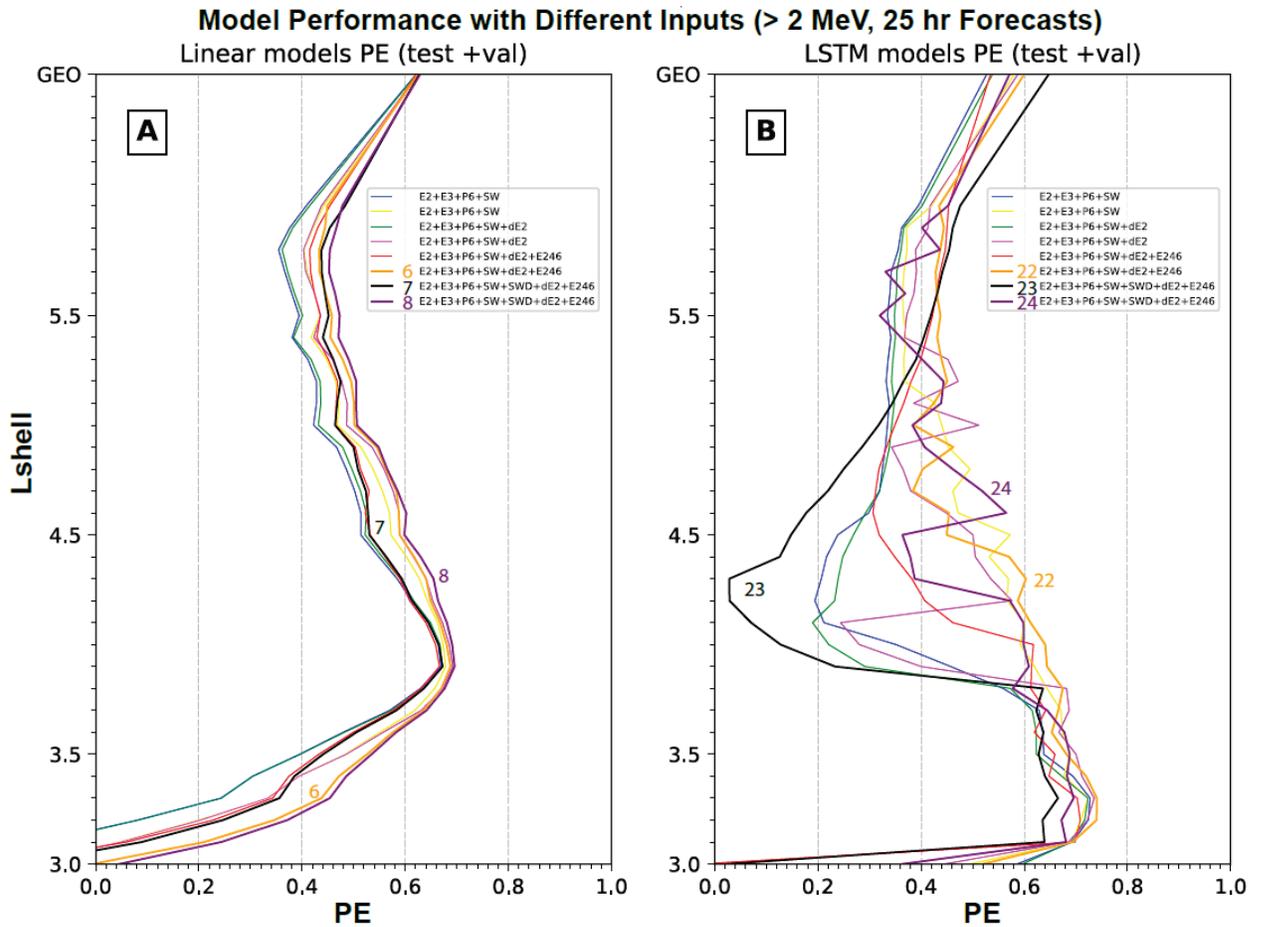

**Figure 2 PE values for the combined validation and test sets are presented as a function of L-shell for linear and LSTM models as in Table 1. A)** Comparison of eight linear regression models with the last three being numbered. **B)** Comparison of eight LSTM models also with the last three being numbered. The models are numbered in the way as in Table 1. Note all linear models behave similarly, but LSTM models vary greatly with different input parameters and window sizes.

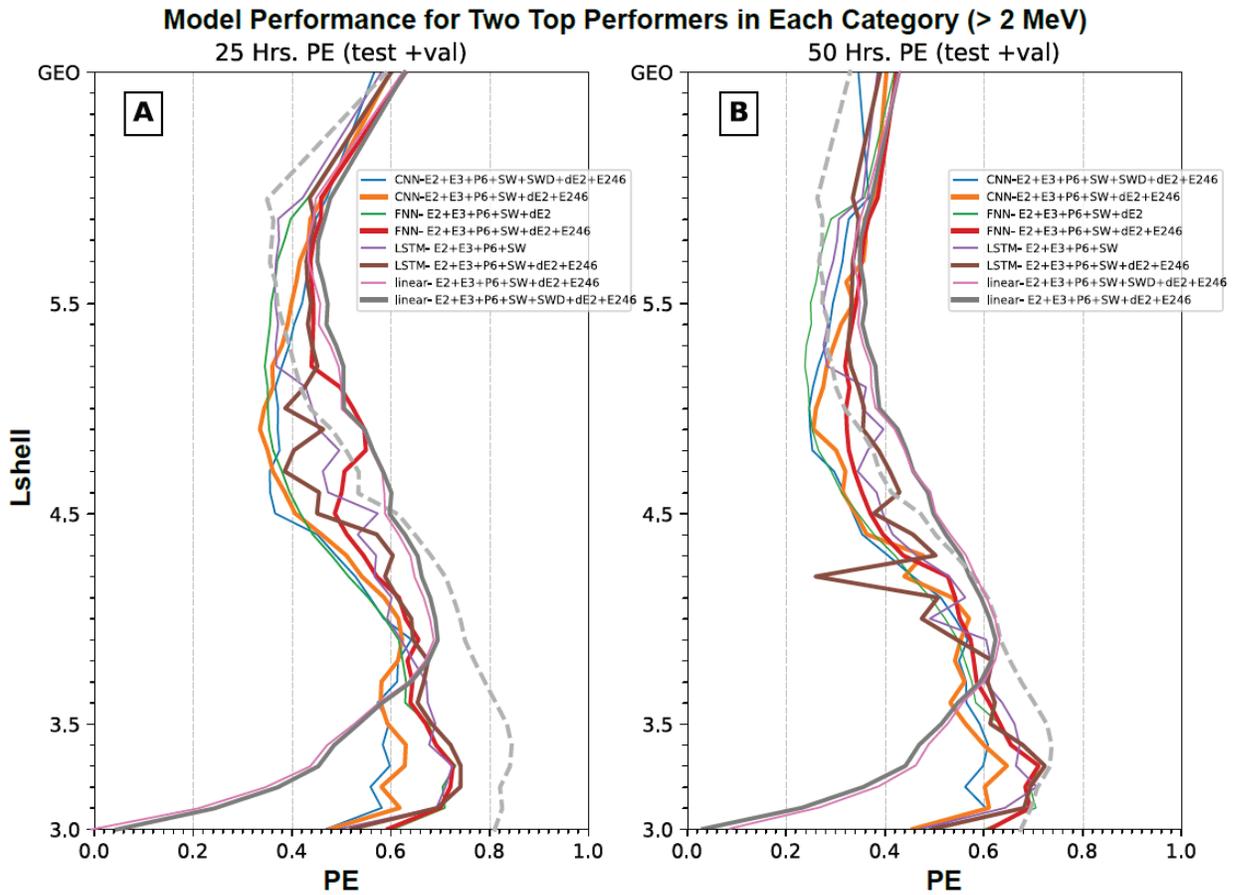

**Figure 3 Model PE values for validation and test data are presented as a function of L-shell for the top two performers in each category forecasting > 2 MeV electrons. A)** Top two performers of each category for 1-day (25 hr) forecasts as listed in Table 1. In the each category, thick (thin) curve is for the top (second) performer. PE curve for the top linear model in PreMevE 2.0 making 1-day forecasts of 1 MeV electrons (P2020) is plotted in dashed gray for comparison. **B)** Top two performers of each category for 2-day (50 hr) forecasts as listed in Table 2. PE curves are in the same format as in Panel A. PE curve for the top linear model in PreMevE 2.0 making 2-day forecasts of 1 MeV electrons (P2020) is also plotted in dashed gray for comparison.

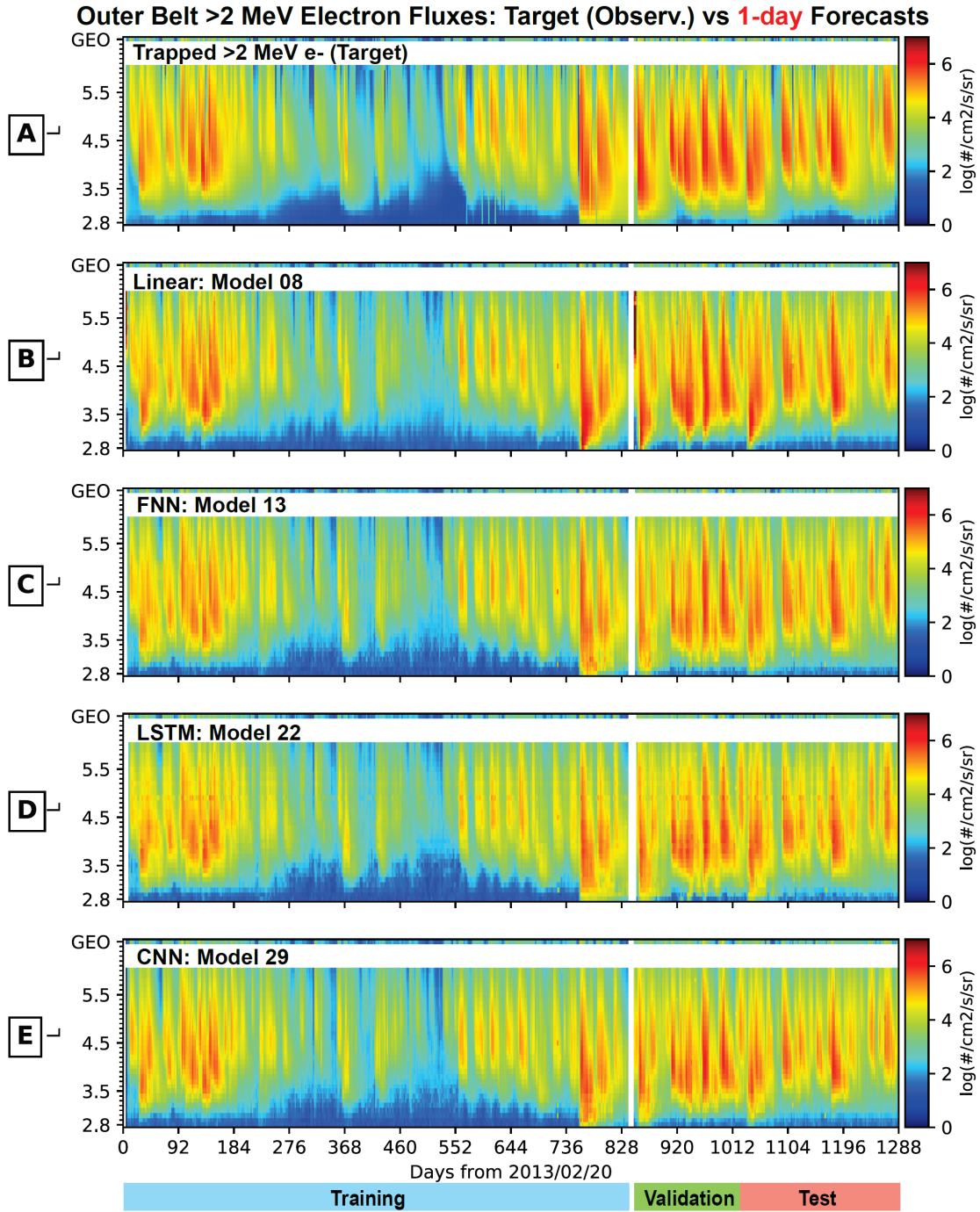

**Figure 4 Overview of target and 1-day forecasted > 2 MeV electron fluxes across all L-shells for the whole 1289-day interval. A**) Observed flux distributions to be forecasted for >2 MeV electrons. Panels **B**) to **E**) show 1-day forecasted flux distributions by the four top performers, each with the highest out-of-sample PE from one category, including the linear regression model 8, FNN model 13, LSTM model 22, and CNN model 29 as listed in Table 1.

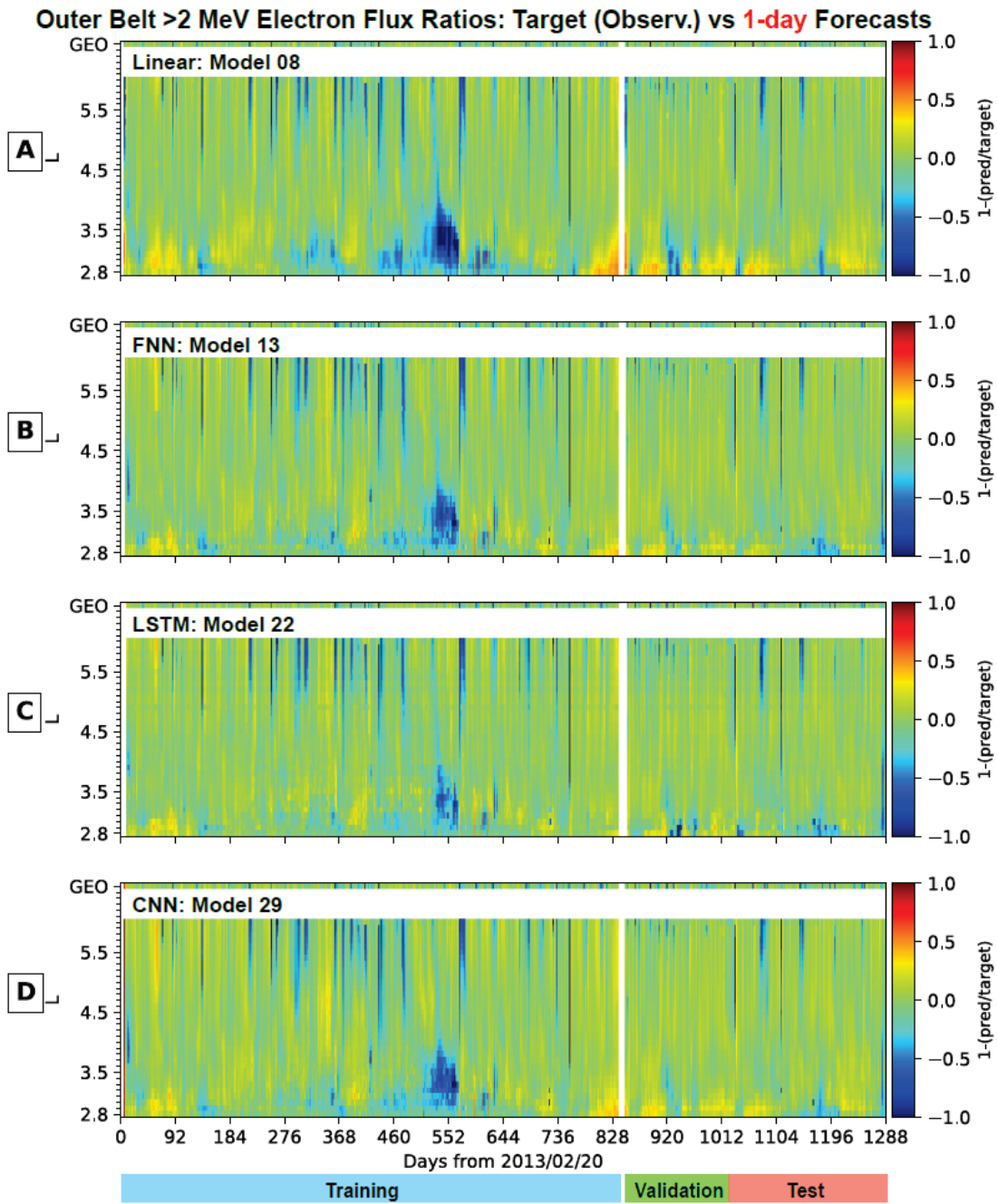

**Figure 5 Relative error ratios of 1-day forecasts across all L-shells for >2 MeV electrons.** Panels A to D plot the deviations ratios, defined as targets minus forecasts and then divided by the targets, as a function of L-shell and time for linear regression model 8, FNN model 13, LSTM model 22, and CNN model 29, respectively, the four top performers as listed in Table 1. Green color depicts perfect predictions, and red (blue) indicates under-predictions (over-predictions).

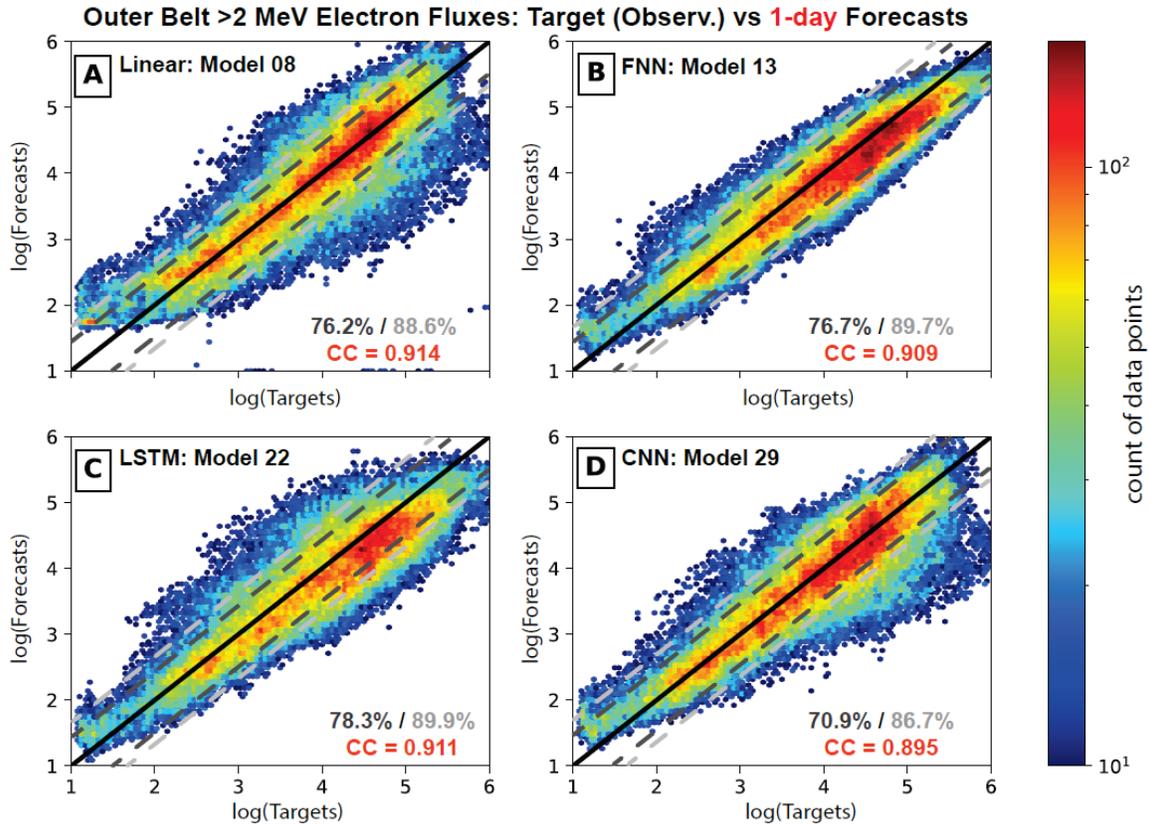

**Figure 6 Model prediction vs. target 2D histograms for 1-day forecasted >2 MeV electron fluxes across all L-shells. A**) Histogram of the fluxes predicted by LinearReg model 8 (the linear top performer as in Table 1) vs. the target >2 MeV electron fluxes. The color bar indicates the count of points in bins of size 0.1 x 0.1. Similarly, panels **B**) to (**D**) show predictions vs >2 MeV target for FNN model 13, LSTM model 22, and CNN model 29, respectively, the top performers as in Table 1. In each panel, diagonal line for perfect matching is shown in solid black curve, and the dashed dark gray (and light gray) lines indicate ratio—between original fluxes—factors of 3 (and 5). The dark gray (light gray) number in lower-right is the percentage of points falling within the factors of 3 (5), and the red number shows the correlation coefficient.

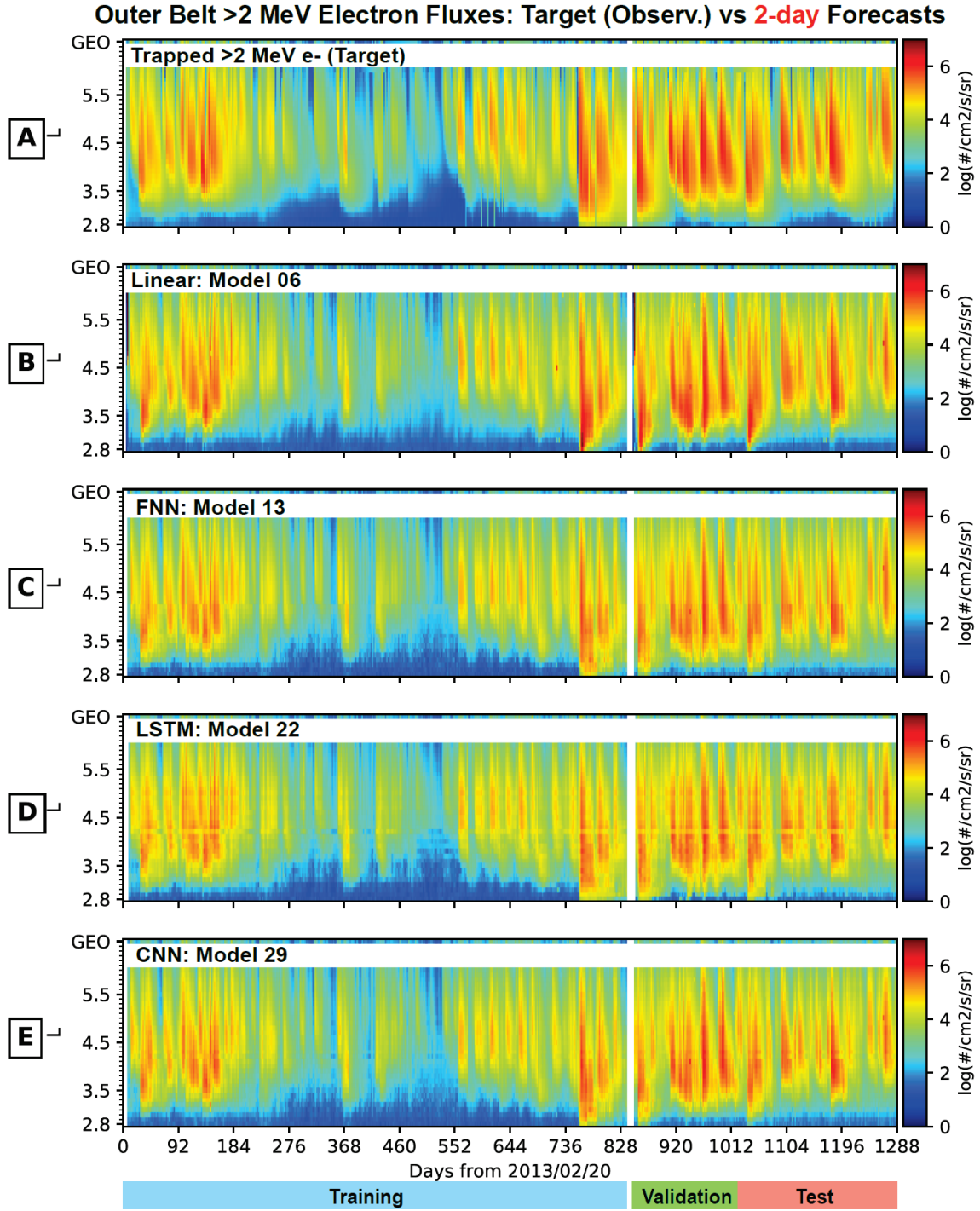

**Figure 7 Overview of target and 2-day forecasted fluxes across all L-shells.** Panel (A) shows the observed flux distributions to be forecasted for >2 MeV electrons. Panels (B) to (E) show, respectively, forecasts from the four top performers, each with the highest out-of-sample PE from one category, including linear regression model 6, FNN model 13, LSTM model 22, and CNN model 29 as listed in Table 2.

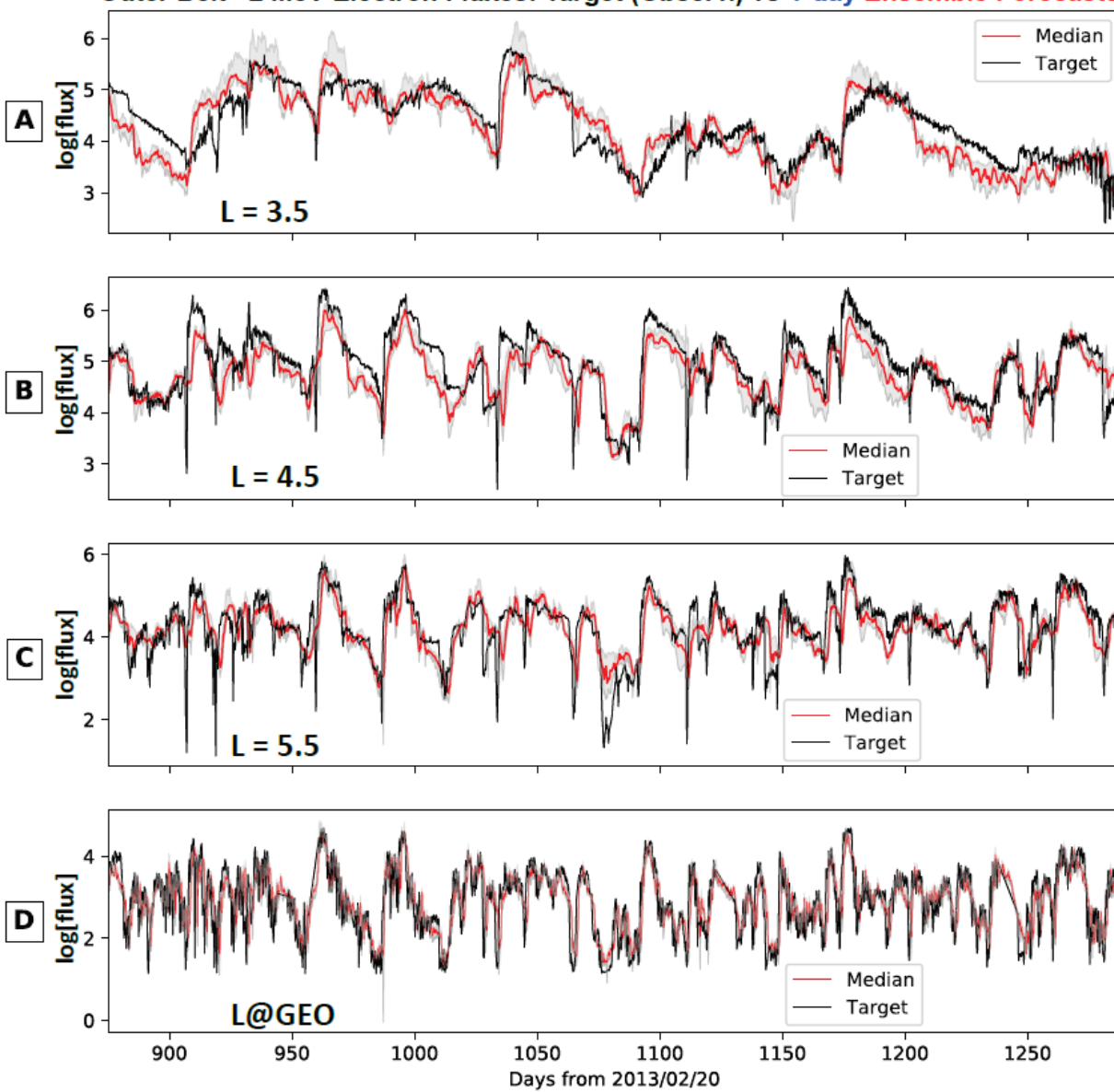

**Figure 8 One-day ensemble forecasting results for > 2 MeV electron fluxes over individual L-shells.** Results are shown for the validation and test periods, and panels from the top to bottom are for L-shells at 3.5, 4.5, 5.5, and GEO (6.6), respectively. In each panel, the target is shown in black, and the gray strip shows the uncertainty ranges (or standard deviations) from the ensemble group, and the median from the ensemble predictions is shown in bright red color. Note that the uncertainties from the ensemble models vary both spatially and temporally, however the median values follow the targets closely.

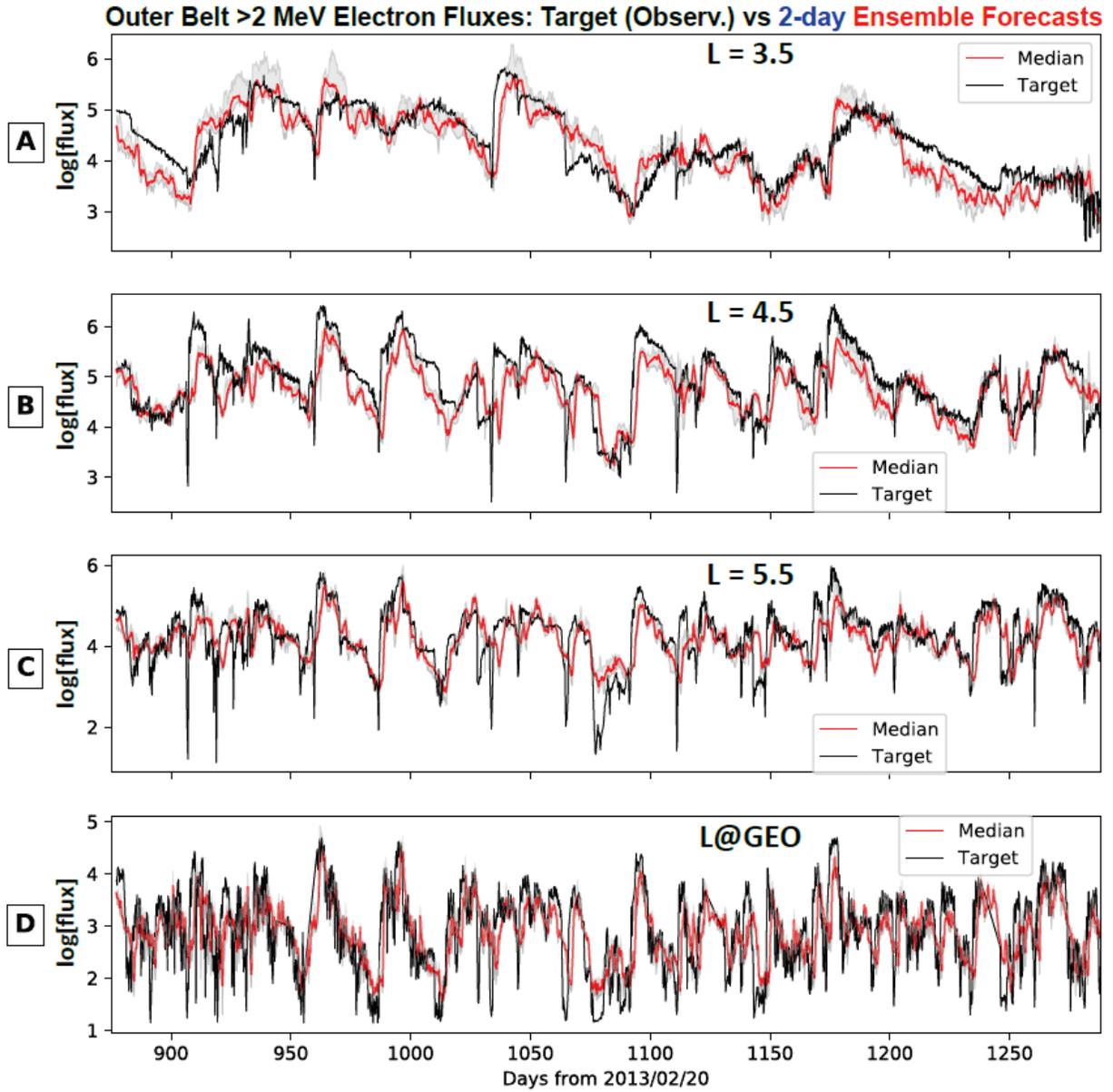

**Figure 9 Two-day ensemble forecasting results for >2 MeV electron fluxes over individual L-shells.** Results are shown for the validation and test periods, and in the same format as Figure 8.

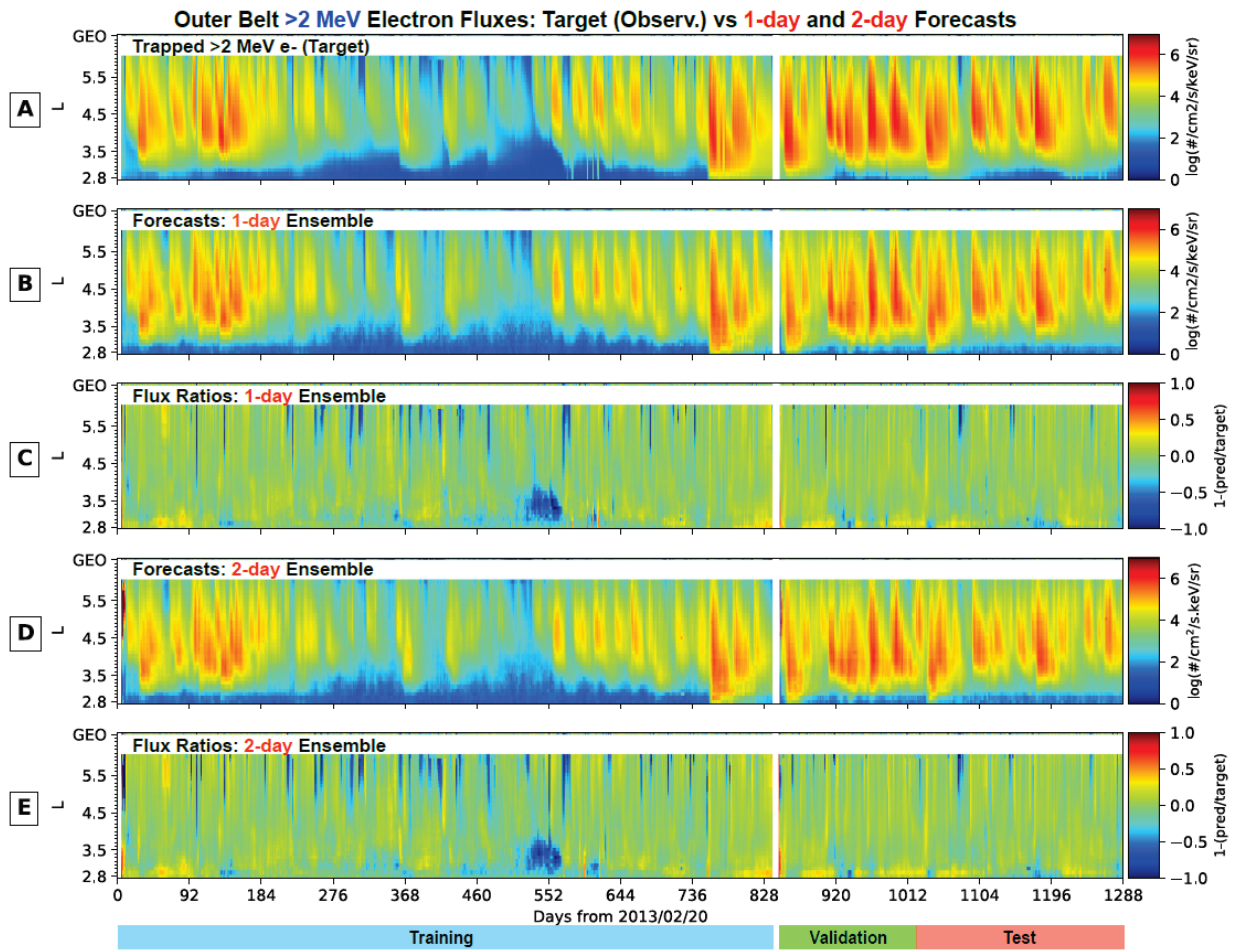

**Figure 10 Overview of target vs 1- and 2-day ensemble forecasted >2 MeV electron fluxes across all L-shells for the whole 1289-day interval. A)** Observed flux distributions. **B)** One-day predicted flux distributions from the ensemble model. **C)** Deviation ratios between the target and 1-day predicted fluxes. **D** and **E)** Same format as B and C but for 2-day ensemble forecasts.

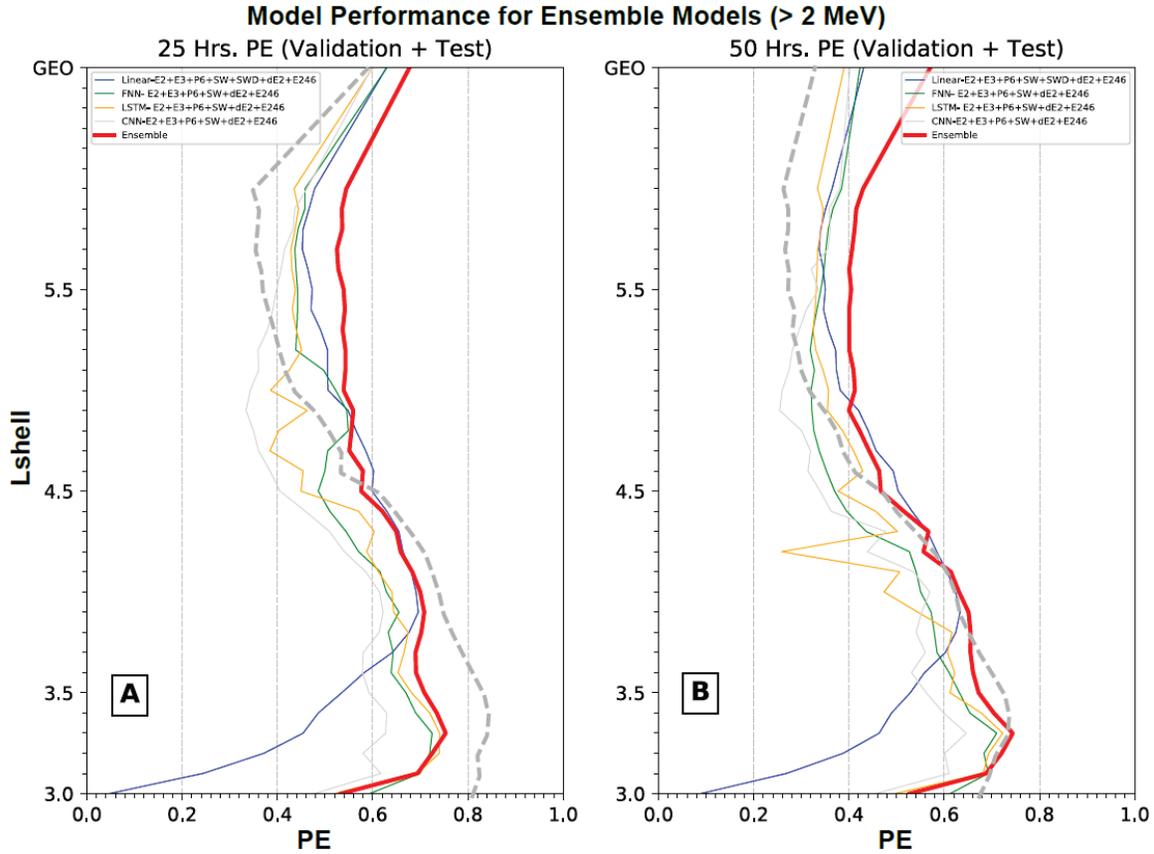

**Figure 11 Model PE values for validation and test data are presented as a function of L-shell for ensemble models forecasting >2 MeV electrons. A)** PE curves for 1-day (25 hr) forecasting models. The thick red curve is for the ensemble model compared to those for four individual ensemble member models (the top performers as defined in Table 1) in different colors. PE curve for the top linear model in PreMevE 2.0 making 1-day forecasts of 1 MeV electrons (P2020) is plotted in dashed gray for comparison. **B)** PE curves for 2-day (50 hr) forecasting models. The red curve is for the ensemble model and other four curves are for ensemble member models (as defined in Table 2). PE curve for the top linear model in PreMevE 2.0 making 2-day forecasts of 1 MeV electrons (P2020) is plotted in dashed gray for comparison.

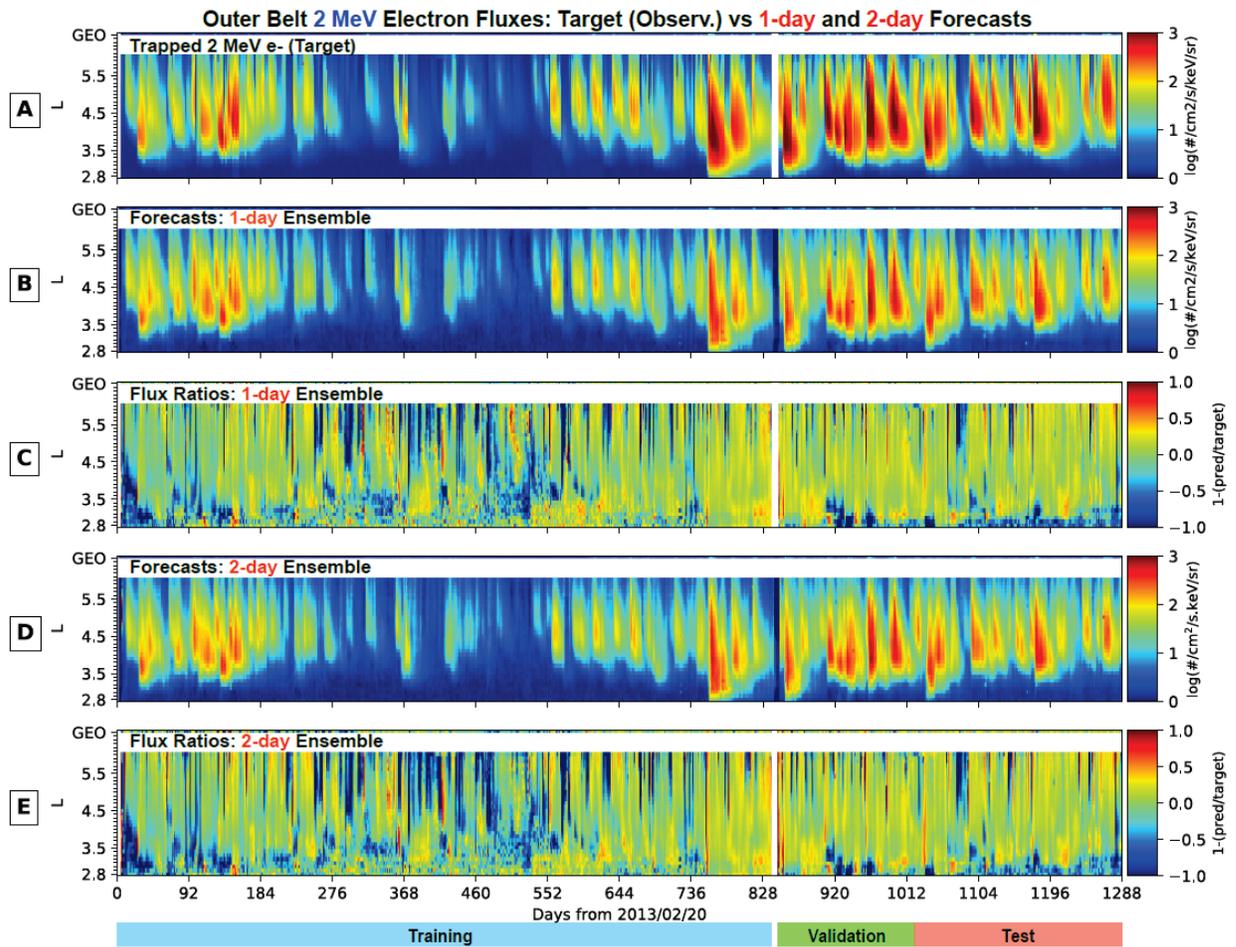

**Figure 12 Overview of target vs 1- and 2-day ensemble forecasted 2 MeV electron fluxes across all L-shells for the whole 1289-day interval.** All panels are in the same format as in Figure 10.

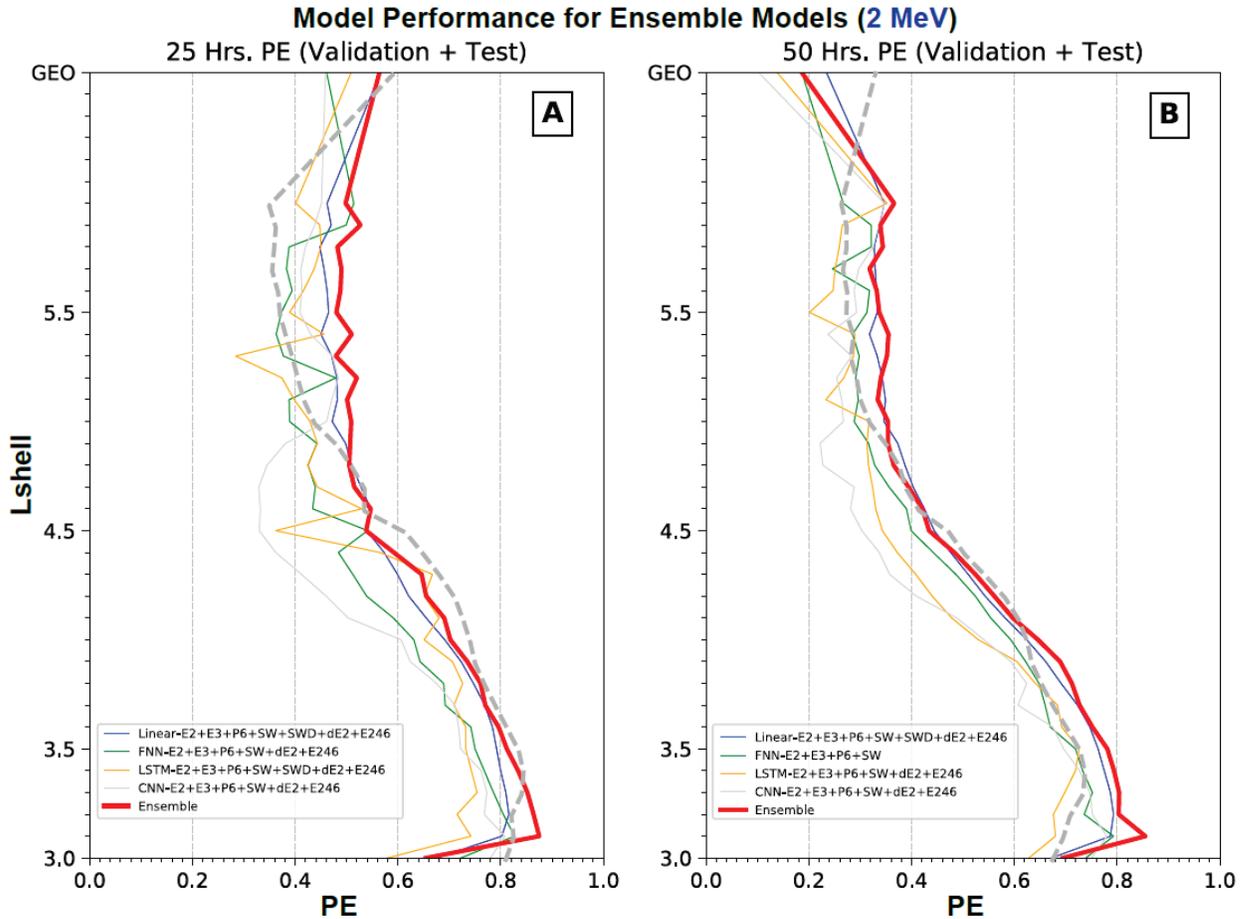

**Figure 13 Model PE values for validation and test data are presented as a function of L-shell for models forecasting 2 MeV electron fluxes. A)** PE curves for 1-day (25 hr) forecasting models. The thick red curve is for the ensemble model (in red) compared to those for four individual ensemble member models (the top performers defined in Table 3) in different colors. **B)** PE curves for 2-day (50 hr) forecasting models. The red curve is for the ensemble model and other four curves are for the ensemble member models (defined in Table 4). PE curves for the top linear model in PreMevE 2.0 making 1- and 2-day forecasts of 1 MeV electrons [Pires de Lima et al., 2020] are plotted for comparison.

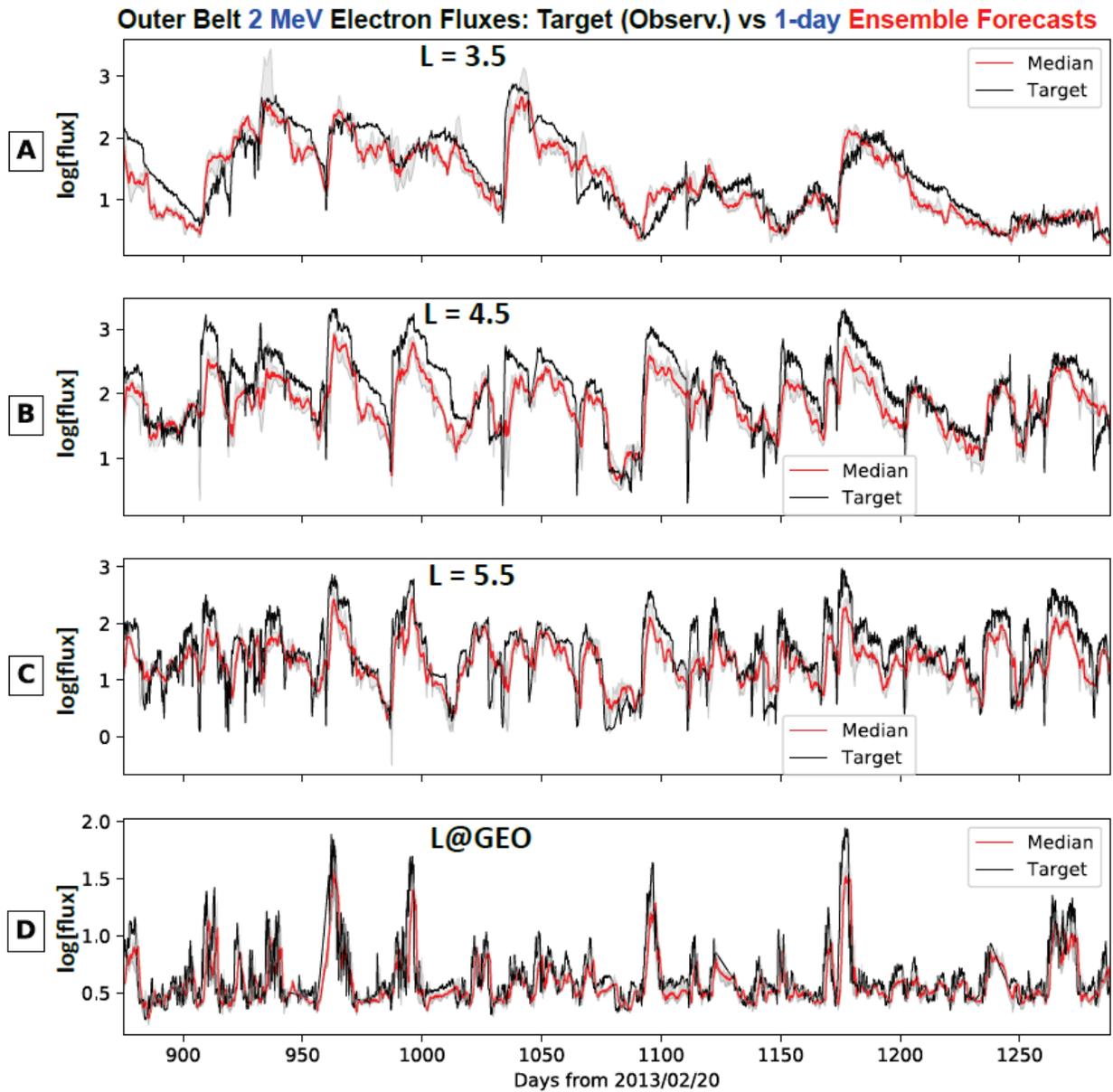

**Figure 14 One-day ensemble forecasting results for 2 MeV electron fluxes over individual L-shells.** Results are shown for the validation and test periods, and panels from the top to bottom are for Lshell at 3.5, 4.5, 5.5, and GEO (6.6), respectively. In each panel, the target is shown in black, and the gray strip shows the uncertainty ranges (or standard deviations) from the ensemble group, and the median from the ensemble predictions is shown in bright red color.

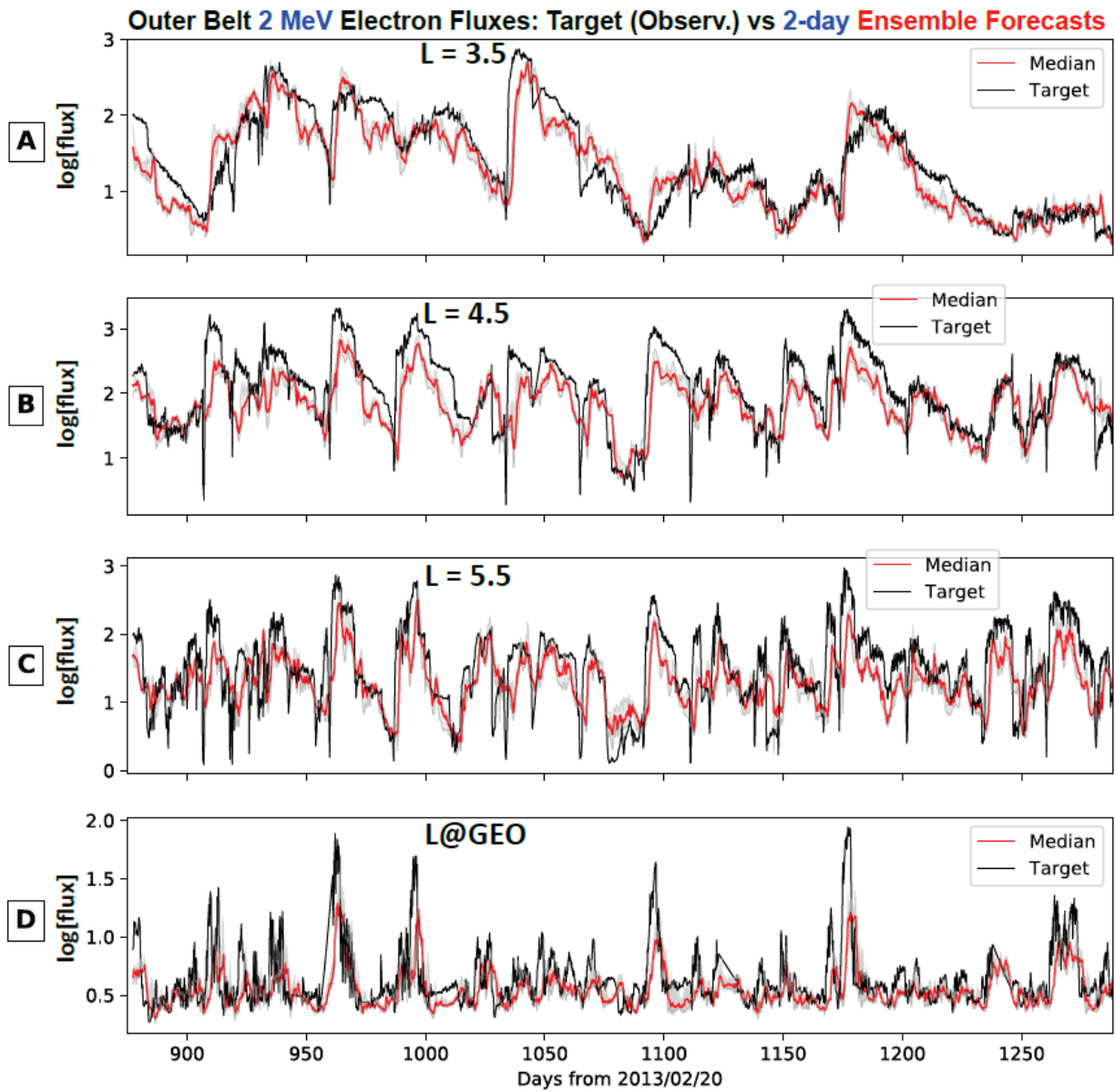

**Figure 15 Two-day ensemble forecasting results for 2 MeV electron fluxes over a range of L-shells.** Results are shown for the validation and test periods, and in the same format as Figure 14.